\documentclass[aps,pre,longbibliography,showpacs,amsmath,amssymb]{revtex4-1}
\usepackage{amsmath,amssymb,latexsym,epsfig,epsf,bm}
\usepackage{graphicx}
\usepackage{float}
\newcommand{\br}{\bold r}

\newcommand{\tr}{\tilde{r}}
\renewcommand{\tt}{\tilde{t}}

\newcommand{\bF}{\bold F}

\newcommand{\bR}{\bold R}

\newcommand{\bRa}{{\bf R}_{\rm a}}

\newcommand{\bRb}{{\bf R}_{\rm b}}

\newcommand{\tbR}{{\tilde{\bold R}}}

\newcommand{\be}{\begin{equation}}
\newcommand{\ee}{\end{equation}}
\newcommand{\fig}[1]{Fig.~\ref{#1}}
\newcommand{\Fig}[1]{Figure~\ref{#1}}
\newcommand{\eq}[1]{Eq.~(\ref{#1})}

\newcommand{\CVex}{{C_V}^{\rm ex}}

\newcommand{\tcVex}{{\tilde{c}_V^{\rm ex}}}

\newcommand{\Sex}{{S}_{\rm ex}}
\newcommand{\tSex}{\tilde{S}_{\rm ex}}

\newcommand{\Eu}{{\rm C}}
\newcommand{\e}{{\rm  EXP}}
\newcommand{\esub}{{\mbox{\tiny EXP}}}

\begin{document}
	\title{The EXP pair-potential system. I. Fluid phase isotherms, isochores, and quasiuniversality}
	\date{\today}
	\author{Andreas Kvist Bacher}\email{ankvba@gmail.com}
	\author{Thomas B. Schr{\o}der}
	\author{Jeppe C. Dyre}\email{dyre@ruc.dk}
	\affiliation{Glass and Time, IMFUFA, Department of Science and Environment, Roskilde University, P.O. Box 260, DK-4000 Roskilde, Denmark}

\begin{abstract}
The exponentially repulsive EXP pair potential defines a system of particles in terms of which simple liquids' quasiuniversality may be explained [A. K. Bacher {\it et al.}, Nat. Commun. {\bf 5}, 5424 (2014); J. C. Dyre, J. Phys. Condens. Matter {\bf 28}, 323001 (2016)]. This paper and its companion present a detailed simulation study of the EXP system. Here we study how structure monitored via the radial distribution function and dynamics monitored via the mean-square displacement as a function of time evolve along the system's isotherms and isochores. The focus is on the gas and liquid phases, which are distinguished pragmatically by the absence or presence of a minimum in the radial distribution function above its first maximum. An $NVU$-based proof of quasiuniversality is presented, and quasiuniversality is illustrated by showing that the structure of the Lennard-Jones system at four selected state points is well approximated by those of EXP pair-potential systems with the same reduced diffusion constant. The companion paper studies the EXP system's isomorphs, focusing also on the gas and liquid phases. 
\end{abstract}
\maketitle

\section{Introduction}\label{I}

For more than half a century the term ``simple liquid'' has implied a system of point particles interacting via pair-wise additive forces \cite{Fisher,ric65,tem68,sti75,bau80,barrat,ing12,han13}. The paradigmatic simple liquid is the hard-sphere (HS) system of identical spheres that do not interact unless they touch each other, at which point the potential energy jumps to infinity \cite{han13,moltheory,ber64,wid67,row73,bar76,cha83}. The HS system embodies a physical picture going back to van der Waal's seminal thesis from 1873 \cite{vdW1873} according to which the harshly repulsive forces between a liquid's atoms or molecules determine the structure. This idea is the basis of the present understanding of liquids as elucidated, e.g., in the classical monograph by Hansen and McDonald from 1976 \cite{han13}, and in the classical reviews by Widom from 1967 \cite{wid67} and by Chandler, Weeks, and Andersen from 1983 \cite{cha83}. The HS picture has had many successes, for instance leading to useful perturbation theories of the liquid state \cite{han13,wca,ros83,kan85,mon01,ben04,zho09,dub14}. 

van der Waals' fundamental insight was that many liquids' properties derive from the repulsive forces \cite{vdW1873}. The weaker and longer-ranged attractive forces play little role for the structure and dynamics; their role is mainly to reduce energy and pressure by providing a virtually constant negative cohesive energy. It has been found from computer simulations, however, that some pair-potential systems are not simple in any reasonable understandings of the term whereas, on the other hand, a number of \textit{molecular} liquids \cite{ing12b} and even polymeric systems \cite{vel14} have simple and regular behavior. A liquid like water exhibits non-simple behavior by having, e.g., a diffusion constant that increases upon isothermal compression, by melting instead of freezing upon compression, etc \cite{jag99}. Pair-potential systems with such anomalous behavior include the Gaussian-core model \cite{sti76,pon09}, the Lennard-Jones Gaussian model \cite{hoa08}, and Jagla model \cite{jag99}. At high and moderate temperatures the Gaussian-core model is not steeply repulsive, which may explain its anomalies, but the other two systems are complex despite their strongly repulsive forces. Thus pair-wise additive forces between point particles are neither necessary nor sufficient for a liquid to be ``simple''. A different definition of simplicity is called for \cite{ing12}.

An alternative definition of liquid simplicity is provided by the isomorph theory according to which simple behavior is found whenever the system in question to a good approximation exhibits ``hidden scale invariance'' (``hidden'' because this property is rarely obvious from the mathematical expression for the potential energy) \cite{IV,ing12,dyr14,dyr16}. This defines the class of Roskilde (R)-simple systems that include the standard Lennard-Jones (LJ) model, a class which was first identified by characteristic strong correlations between the virial and potential-energy thermal fluctuations in the canonical ($NVT$) ensemble \cite{ped08,ped08a,I}. 

R-simple systems have isomorphs \cite{IV}, which are lines in the thermodynamic phase diagram along which structure and dynamics in reduced units (see below) are invariant to a good approximation. These invariances reflect the fact that state points on the same isomorph have approximately the same canonical probabilities for configurations that scale uniformly into one another \cite{IV}. Isomorph-theory predictions have been validated in computer simulations of Lennard-Jones type systems \cite{IV,V,cos16a}, simple molecular models \cite{ing12b}, crystals \cite{alb14}, nano-confined liquids \cite{ing13a}, non-linear shear flows \cite{sep13}, zero-temperature plastic flows of glasses \cite{ler14}, polymer-like flexible molecules \cite{vel14,vel15a}, metals studied by \textit{ab initio} density functional theory computer simulations \cite{hum15}, plasmas \cite{vel15}, and other simple liquids \cite{dyr16,bac14a}. Experimental confirmations of the isomorph theory were presented in Refs. \onlinecite{nie08,gun11,roe13,xia15,nis17,han18}. The numerical and experimental confirmations notwithstanding, it is important to emphasize that the isomorph theory is rarely exact, that it usually works only in the liquid and solid parts of the thermodynamic phase diagram (the EXP system is an exception to this), and that the theory does not apply for systems with strong directional bonding (hydrogen-bonding or covalently bonded systems). 

The basic characteristic of an R-simple system is that, because of its isomorphs, the thermodynamic phase diagram is effectively one-dimensional \cite{IV,dyr16}. Notably, R-simple systems have this property in common with the HS system for which the packing fraction determines the physics throughout the phase diagram \cite{han13}.

In 2014 it was shown \cite{sch14} that the isomorph theory is a consequence of the following scale-invariance property in which $\bR=(\br_1,...,\br_N)$ is the vector of all particle coordinates, $U(\bR)$ the potential-energy function, and $\lambda$ a uniform scaling parameter:

\be\label{hs}
U(\bRa)<U(\bRb)\,\Rightarrow\, U(\lambda\bRa)<U(\lambda\bRb)\,.
\ee
Thus if the potential energy of some configuration $\bRa$ is lower than that of another configuration $\bRb$, both of same density, this property is maintained after a uniform scaling of the configurations. Strong virial potential-energy correlations \cite{ped08}, as well as the approximate invariance along isomorphs of Boltzmann probabilities of uniformly scaled configurations that originally defined isomorphs \cite{IV}, are consequences of \eq{hs} \cite{sch14}. The scale-invariance property \eq{hs} is only obeyed rigorously for the unrealistic case of a system with an Euler-homogeneous potential-energy function plus a constant. For realistic R-simple systems \eq{hs} applies to a good approximation, i.e., for modest density variations of most of its physically relevant configurations. This is, nevertheless, enough to ensure approximate invariance of structure and dynamics along the isomorphs \cite{sch14}. Incidentally, these curves in the thermodynamic phase diagram are virtually parallel to the freezing and melting lines \cite{IV,bac14a}, a fact that explains several well-known phenomenological melting-line characterizations like, e.g., the Lindemann melting criterion \cite{IV,cos16,ped16}.

The invariance of structure and dynamics along isomorphs relates to ``reduced'' quantities \cite{IV,sch14,dyr16}. These are quantities that have been made dimensionless by scaling with the \textit{length} 

\be
l_0\equiv\rho^{-1/3}
\ee
defined from the particle density $\rho\equiv N/V$ in which $N$ is the number of particles and $V$ is the sample volume, the \textit{energy} 

\be
e_0\equiv k_BT
\ee
in which $T$ is the temperature, and the \textit{time} 

\be
t_0\equiv\rho^{-1/3}\sqrt{\frac{m}{k_BT}}
\ee
in which $m$ is the average particle mass. Note that these units vary with the state point in question. Reduced units are used throughout the present paper and its companion \cite{EXP_II_arXiv}. Two notable exceptions to this are density and temperature, which are both constant in reduced units. Therefore, in order to specify a state point, density is reported in units of the EXP pair potential length parameter $\sigma$ of \eq{EXPppX} below, i.e., in units of $1/\sigma^3$, and temperature is reported in units of the potential's energy parameter, $\varepsilon/k_B$. We refer to this as the ``EXP unit system''.

Although the HS system provides a good reference for understanding simple liquids, it has some challenges \cite{bac14a}. For instance, while simple liquids' quasiuniversal \textit{structure} may be understood from the harsh interparticle repulsions modeled by a HS system, it is much less obvious how to explain simple liquids' quasiuniversal \textit{dynamics} by reference to the HS system. After all, the HS system's particles evolve in time following straight lines in space interrupted by infinitely fast collisions, which is quite different from what happens in a real liquid where each particle interacts continuously and strongly with ten or more nearest neighbors. Also, the HS reference system cannot explain the above-mentioned fact that some systems with strong interparticle repulsions do not belong to the quasi-universal class of ``simple'' systems \cite{dyr16}. Finally, the HS system is unphysical by having a discontinuous potential-energy function, implying in particular that the time-averaged potential energy is zero at all state points.

It would be nice to have a generic analytic pair-potential system in terms of which simple liquids' quasiuniversality may be explained, defining the ``mother of all pair-potential systems''. It was recently suggested \cite{bac14a,dyr16} that this role may be played by the exponentially repulsive EXP pair potential defined by (in which $\varepsilon$ is a characteristic energy and $\sigma$ a characteristic length)

\be\label{EXPppX}
v_\e(r)\,=\,\varepsilon\, e^{-r/\sigma}\,.
\ee
Refs. \onlinecite{bac14a,dyr16} showed that any system with a pair potential, which may be written as a sum of spatially decaying exponentials of the form given in \eq{EXPppX} with numerically large prefactors relative to $k_BT$, to a good approximation obeys the same equation of motion as the EXP system itself. This explains the quasiuniversality of traditional simple liquids like the Lennard-Jones system, inverse power-law systems, Yukawa pair potential, etc, as well as exceptions to quasiuniversality that cannot be written in this way \cite{bac14a,dyr16}.

	\begin{figure}[!htbp]
	\centering
	\includegraphics[width=0.45\textwidth]{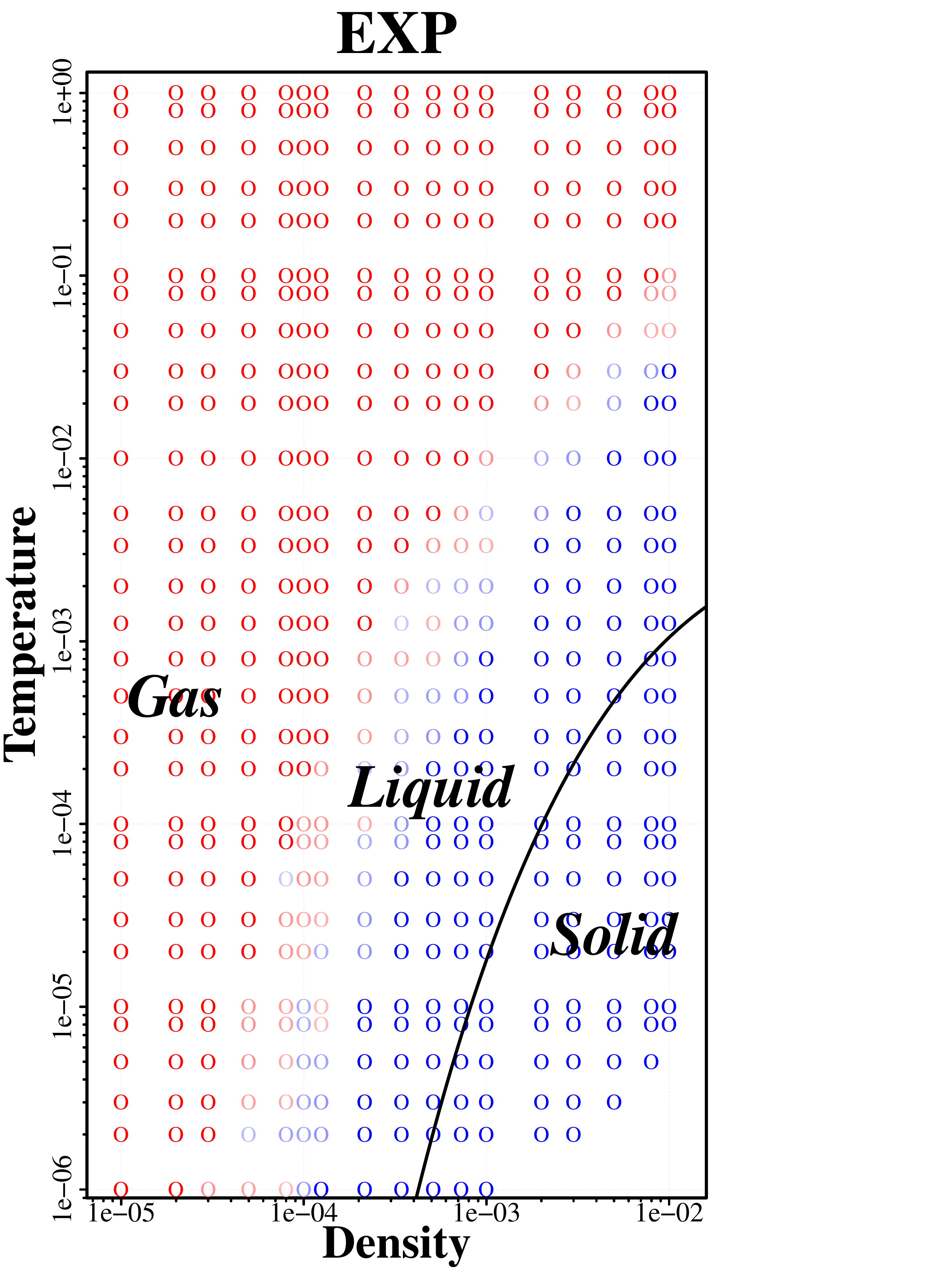}
	\caption{\label{Fig1}Log-log density-temperature phase diagram of the EXP system showing the state points investigated. Gas-phase state points are given in red, condensed-phase (liquid and solid) state points in blue. Because the EXP pair potential is purely repulsive, there is no gas-liquid phase transition and no gas-liquid coexistence region; the gas and liquid phases merge continuously. Liquid state points are distinguished pragmatically from those of the gas phase by having minima in their pair-distribution functions above the nearest-neighbor peak; the light colored state points indicate the transition region between gas and liquid. The solid-liquid coexistence region is covered by the black line (determined as the approximate melting-line isomorph).}
\end{figure}

Despite its mathematical simplicity and the fact that the exponential function in mathematics is absolutely central, e.g., for defining the Fourier and Laplace transforms, the EXP pair-potential system has been studied little on its own right. In the literature an EXP term typically appears added to an $r^{-6}$ attractive term \cite{bor32,buc38} or multiplied by a $1/r$ term as in the Yukawa pair potential \cite{yuk35,row89}. Born and Meyer in 1932 used an exponentially repulsive term in a pair potential and justified this from the fact that electronic bound-state wavefunctions decay exponentially in space \cite{bor32}. Kac and coworkers used a HS pair potential minus a long-ranged EXP term for rigorously deriving the van der Waals equation of state in one dimension \cite{kac63}. Recently, by reference to the EXP pair potential Maimbourg and Kurchan showed that the isomorph theory for pair-potential systems with strong repulsions becomes exact in infinite dimensions \cite{mai16}. The EXP pair potential was also used recently by Kooij and Lerner in as study of unjamming in models with analytic pair potentials \cite{koo17}. 

The reason the pure EXP pair-potential system has not been studied very much may be that this system has been regarded as unrealistic by being purely repulsive. However, even the purely repulsive inverse power law (IPL) pair-potential systems have been studied much more than the EXP system \cite{hiw74,hey07,hey08,lan09,bra11,pie14,din15}. In view of this, the present paper undertakes an investigation of the EXP pair-potential system by presenting results from extensive computer simulations. 

\Fig{Fig1} shows the phase diagram of the EXP system indicating the state points studied in the present paper. Like for any purely repulsive system there are only two thermodynamically distinct phases: a solid phase at low temperatures and high densities and a ``fluid'' phase; there is no gas-liquid phase transition since this requires attractive forces. We have chosen nevertheless to pragmatically distinguish typical ``gas'' state points from typical ``liquid'' state points, but it is important to recall throughout the paper that these phases merge continuously into one another, just as in a real system above its critical temperature. The reason we distinguish between gas and liquid state points is the significance of quasiuniversality (Sec. \ref{QUA}), which implies  that close to the melting line the EXP system's behavior in regard to both structure and dynamics is indistinguishable from that of, e.g., the Lennard-Jones liquid close to freezing. To distinguish the gas and liquid phases we used the following criterion: if the radial distribution has a clear minimum above its first maximum, the state point is liquid, if not it is a gas-phase state point. There is a large region of in-between states, which is indicated in \fig{Fig1} by the use of light colors.

In Sec. \ref{II} we briefly discuss technicalities relating to computer simulations of the EXP system. Section \ref{III} shows that the EXP system obeys \eq{hs} to a good approximation by demonstrating that one of its consequences -- strong virial potential-energy correlations at constant density \cite{I,II} -- applies in a large part of the thermodynamic phase diagram. Section \ref{IIIa} gives results for how pressure and average potential energy vary throughout the system's thermodynamic phase diagram. In Sec. \ref{IV} we report simulations of structure and dynamics along isotherms, while Sec. \ref{V} gives the same information along isochores. Even though the EXP system has no liquid-gas phase transition, its structure and dynamics look pretty much like those of other simple liquids. Section \ref{QUA} rationalizes this by proving quasiuniversality in terms of the EXP pair-potential system. Here we also present numerical results for four state points, showing that the physics of the Lennard-Jones system is fitted well by that of EXP systems with the same reduced diffusion constant. Finally, Sec. \ref{VI} provides a brief summary.

\begin{figure}[!htbp]
		\centering
		\includegraphics[width=0.45\textwidth]{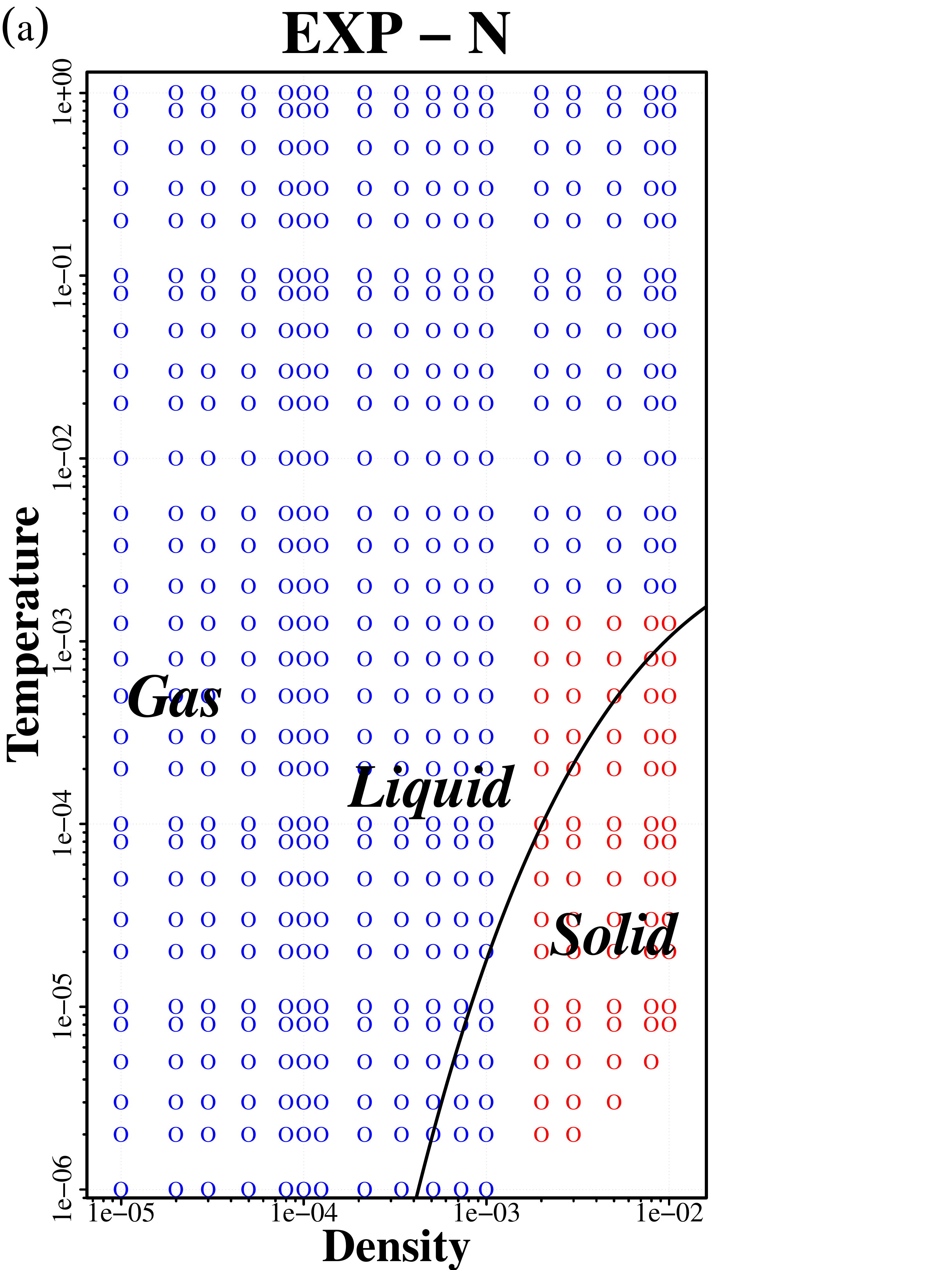}
		\includegraphics[width=0.45\textwidth]{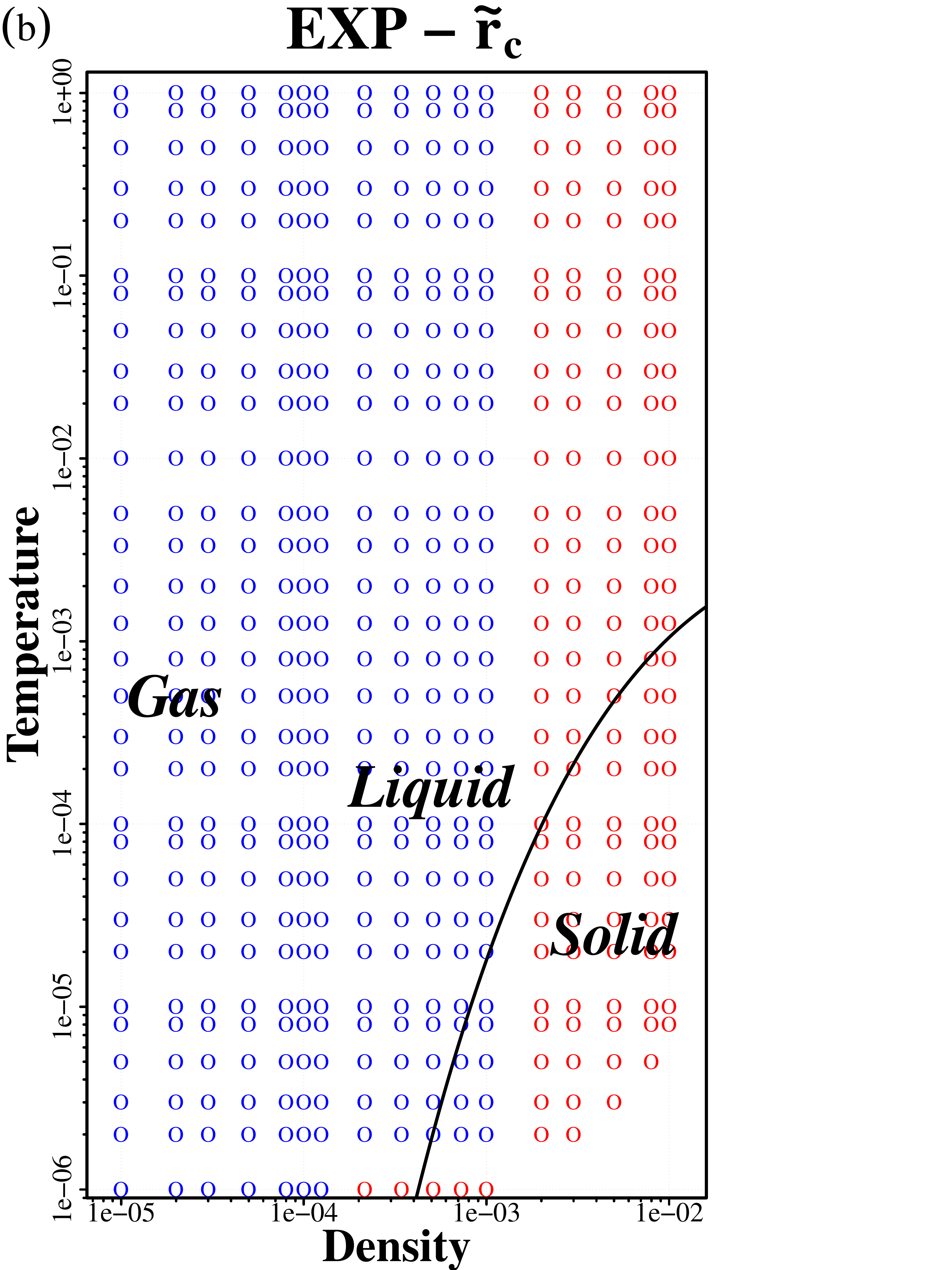}
		\caption{(a) Number of particles $N$ simulated at different state points where blue indicates 1000 particles and red 2000 particles. (b) Shifted-force cutoff $r_c$ in reduced units ($\tilde{r}_c\equiv r_c\rho^{1/3}$) where blue indicates $\tilde{r}_c=2$ and red $\tilde{r}_c=4$.\label{fi:simdat}}
\end{figure}

\section{Simulations details}\label{II}

The simulations were performed on graphics cards using the RUMD Molecular Dynamics open-source software \cite{rumd15}. All simulations were carried out using the unit system in which temperature and density are both equal to one; varying the state point is thus achieved by changing the parameters $\varepsilon$ and $\sigma$ of the EXP pair potential.

The time step $\Delta t = 0.0025$ was used in most of the phase diagram, except for state points with $T=10^{-6}$ and $\rho>2\cdot 10^{-4}$ for which $\Delta t = 0.002$. Temperature was controlled by a Nose-Hoover thermostat with characteristic time 0.2. For most state points an initial configuration of 1000 particles in a simple cubic lattice was generated with thermal velocities. An initial configuration of 2000 particles in a body-centered cubic lattice was used for state points with $1.5\cdot 10^{-6}<T<1.5\cdot 10^{-3}$ and $\rho>1.5 \cdot 10^{-3}$, compare \fig{fi:simdat}(a). The initial configuration was equilibrated by 10,000,000 time steps at the desired state point - ensuring a mean-square displacement of at least 1000 at all fluid state points. Data collection was carried out over at least 10,000,000 subsequent time steps.
 
A shifted-force cutoff was used to allow for a shorter cutoff distance than the standard shifted-potenial cutoff \cite{tildesley,tox11a}. The cut-off was 2$\sigma$ when $\rho<1.5 \cdot 10^{-3}$ except for the lowest-temperature state points, else at 4$\sigma$, compare \fig{fi:simdat}(b). RUMD uses single precision as standard. A customized version with double precision was used to validate selected simulations, concluding that single precision works well in the reported part of the phase diagram. Only in the low-temperature, high-density part, i.e., deep into the crystalline phase, did the use of single precision present a problem. These state points have been left out. 

The focus of the present paper and the companion paper \cite{EXP_II_arXiv} is on the gas and liquid phases. Results are occasionally reported also for the solid (crystalline) phase, but they may be less reliable by deriving from simulations initiated from lattices that in some cases during the simulation reorganized into different crystal structures. This led to crystals with many defects, i.e., solids that are not in proper thermodynamic equilibrium.

\section{Strong virial potential-energy correlations}\label{III}

This section studies how well the EXP system's constant-density thermal-equilibrium virial fluctuations correlate with its potential-energy fluctuations. Strong such correlations are a consequence of \eq{hs} \cite{sch14} and have been demonstrated in $NVT$ computer simulations of many model liquids \cite{ped08,I}, including molecular ones \cite{sch09}. Recall that the microscopic virial $W(\bR)$ is defined as $W(\bR)=\partial U(\bR)/\partial\ln\rho$ in which it is assumed that the density change induces a uniform scaling of $\bR$. The virial, which is an extensive quantity of dimension energy, provides the modification of the ideal-gas law caused by particle interactions: 

\be\label{EOS}
pV=Nk_BT+W
\ee
in which $W=\langle W(\bR)\rangle$ where the sharp brackets denote a thermal average.

	\begin{figure}[!htbp]
	\includegraphics[width=0.45\textwidth]{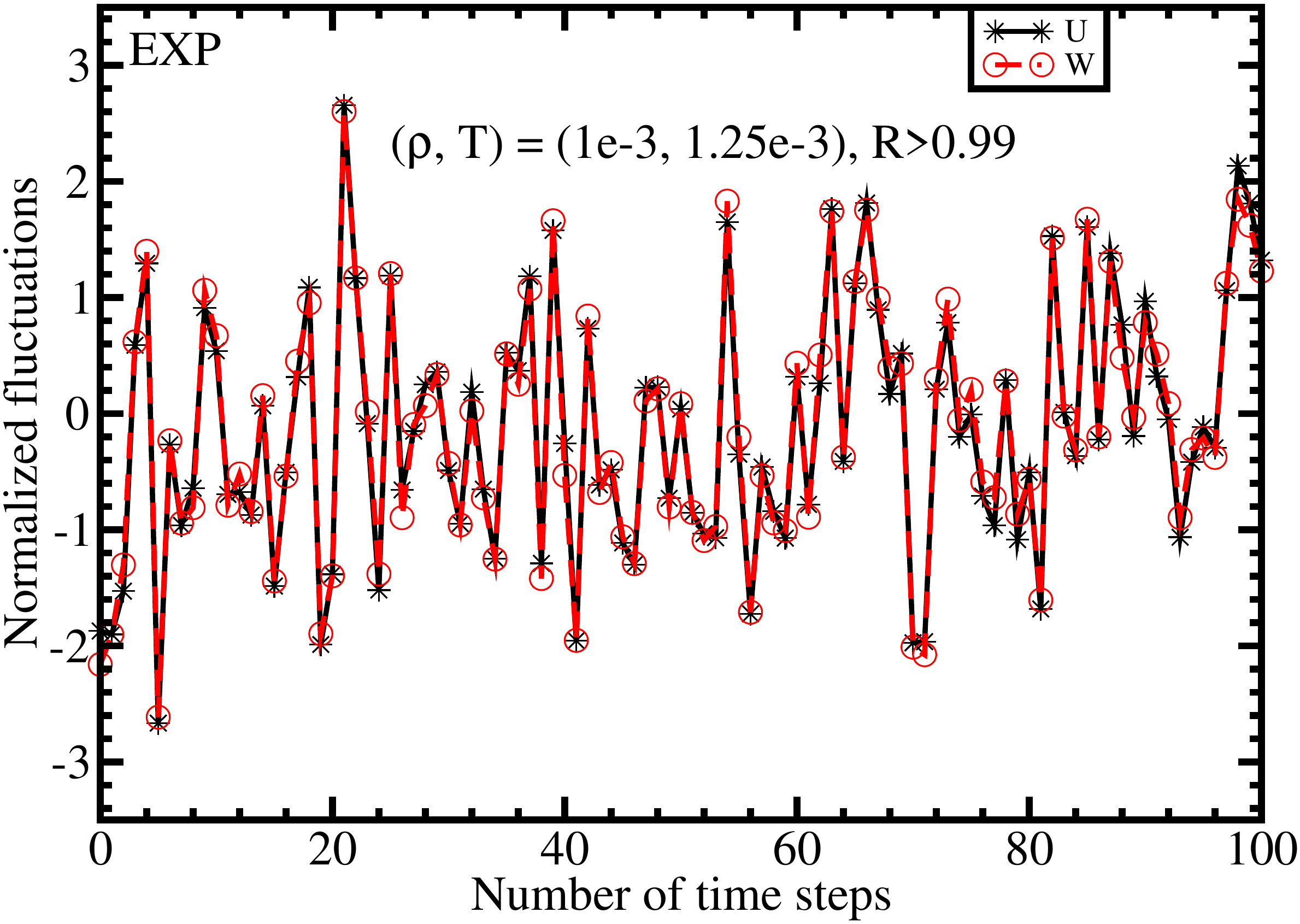}
	\caption{Normalized equilibrium fluctuations of potential energy (black) and virial (red) at the state point $(\rho,T)=(10^{-3},1.25\cdot 10^{-3})$. The correlations are strong ($R=0.9916$), showing that the EXP pair potential system is R-simple at this state point.
	\label{fi:dUdW-scatter}}
\end{figure}

The microscopic virial is calculated by summing over all particles as follows: $W(\bR)=\sum_{i<j} \br_{ij}\cdot\bF_{ij}/3$ where $\br_{ij}$ is the position vector from particle $i$ to particle $j$ and $\bF_{ij}$ is the force with which particle $i$ acts on particle $j$ \cite{tildesley,han13}. \Fig{fi:dUdW-scatter} shows results from simulations of the EXP system's equilibrium fluctuations at a liquid state point. The black stars give the potential energy and the red circles give the virial. Both quantities have been subtracted the mean and normalized to unit variance. There is a very strong correlation. This is perhaps not surprising since the EXP pair potential has the unique property that the pair force is proportional to the pair potential energy. Thus, if interactions corresponding to a narrow range of pair distances dominate the potential energy as well as the virial, one expects strong correlations between these two quantities.

The Pearson correlation coefficient $R$ quantifying correlations is defined by

\be\label{R_def}
R
\,\equiv\,\frac{\langle\Delta U\Delta W\rangle}{\sqrt{\langle(\Delta U)^2\rangle\langle(\Delta W)^2\rangle}}\,
\ee
in which $\Delta$ denotes the quantity in question minus its state-point average. A system is defined to be R-simple or strongly correlating whenever $R>0.9$ \cite{I}, which provides a pragmatic though somewhat arbitrary criterion. The state point studied in \fig{fi:dUdW-scatter} has better than 99\% correlation. This is quite strong compared to, for instance, the R-simple Lennard-Jones system that has $R\sim 95$\% for liquid state points close to the triple point.

\begin{figure}[!htbp]
		\centering
		\includegraphics[width=0.45\textwidth]{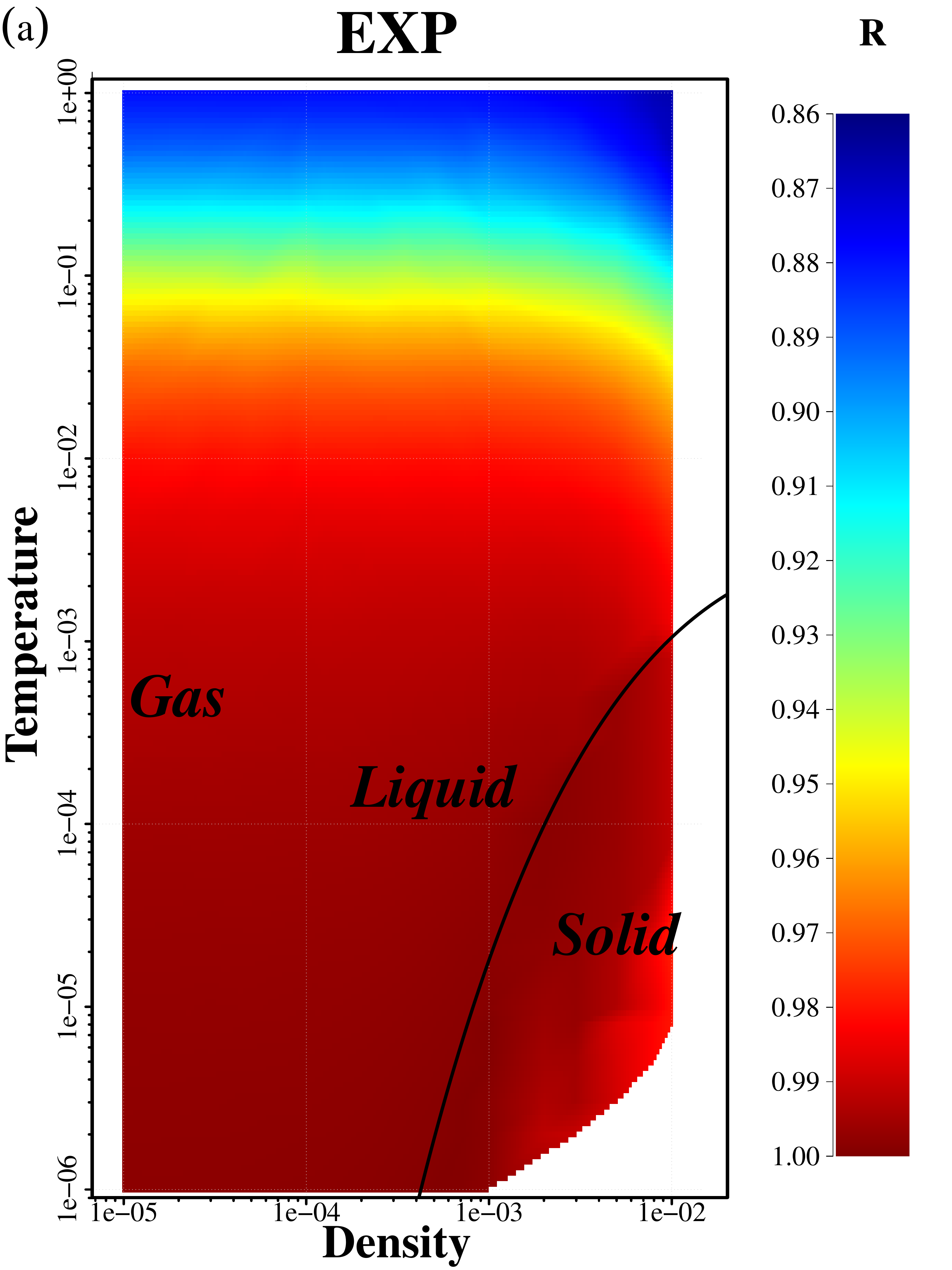}
		\includegraphics[width=0.45\textwidth]{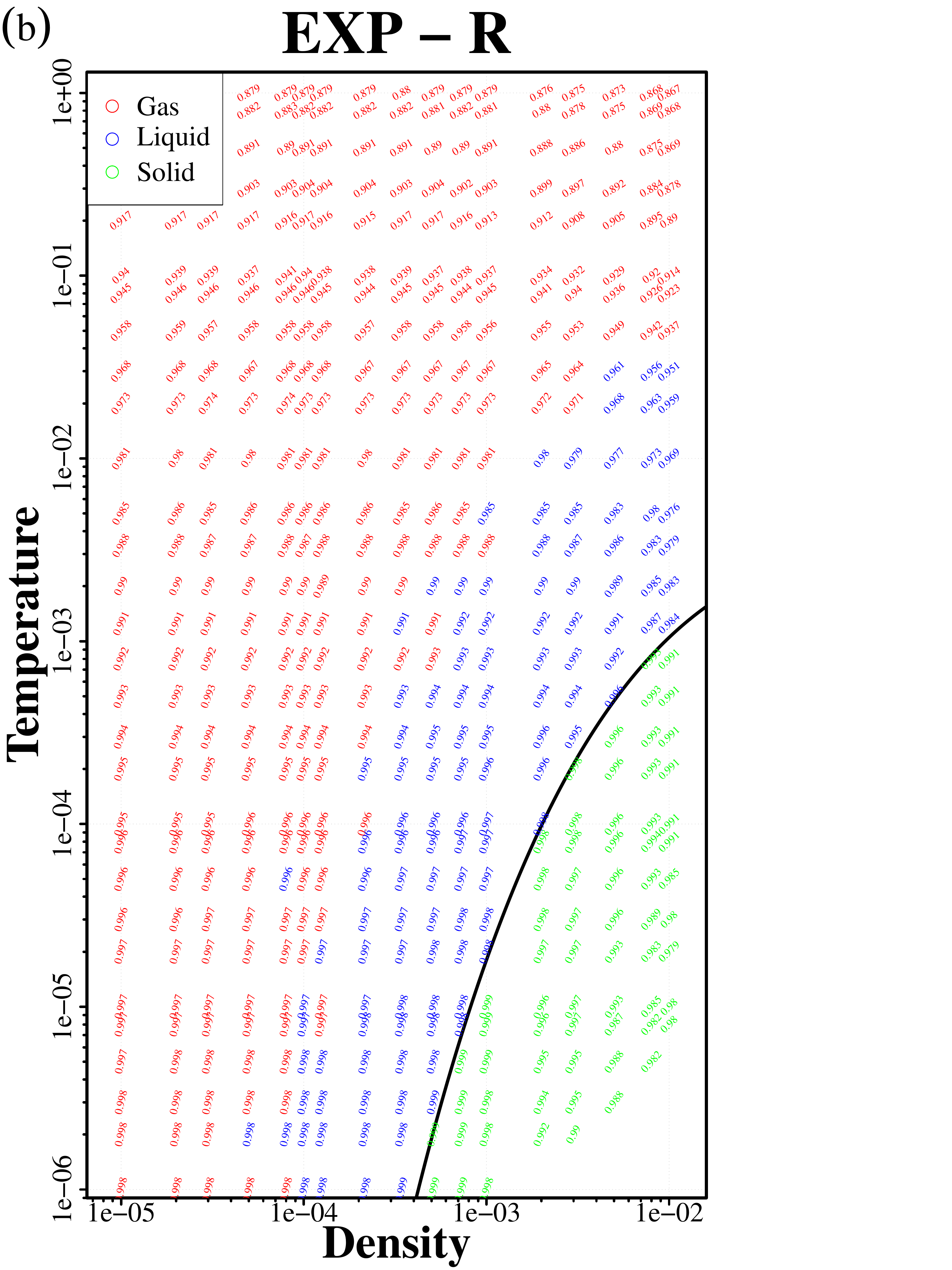}
		\caption{\label{fi:phase-diagram}(a) Phase diagram of the EXP system giving the virial potential-energy correlation coefficient $R$ (\eq{R_def}) in color coding. The EXP system is R-simple in the low-temperature part of the phase diagram. The correlation coefficient depends predominantly on the temperature, which is the prediction of the analytical theory for $R$ in the gas phase (Appendix I). The black line covers the solid-liquid coexistence region.
		(b) Numerical values of $R$ at different state points (visible upon magnification) at the densities and temperatures listed in Appendix II. At each state point the value of $R$ is written with a slope marking the direction of the isomorph through the state point in question. Red indicates gas, blue liquid, and green solid phase state points.}\end{figure}

We evaluated $R$ for the EXP system at several state points. \Fig{fi:phase-diagram}(a) shows the thermodynamic phase diagram colored after the value of $R$, while (b) gives numerical values of $R$ throughout the phase diagram (tiny numbers written into the figure). Interestingly, $R$ is fairly independent of the density. The virial potential-energy correlation coefficient is close to unity at low temperatures (Paper II argues that $R\rightarrow 1$ as $T\rightarrow 0$ if this limit is taken along an isomorph). Note that $R\cong 1$ applies at low temperatures for all phases, which may be interpreted as reflecting an effective inverse-power law behavior of the EXP system at low temperatures, independent of density. In particular, it is notable that the low-temperature gas phase exhibits strong correlations; this may be contrasted to the Lennard-Jones system for which this does not apply due to the attractive pair forces \cite{I,II}.

\begin{table} 
	\begin{tabular}{c|c|c}
	{$\,\,$\rm Temperature} & $\,\,$$R$ {\rm from theory } &$\,\,$ $R$ {\rm from gas-phase simulations }\\ %$
	\hline 
	$1.00\cdot 10^{-1}$  & 0.9396 & 0.9348 \\   
	$1.00\cdot 10^{-2}$  & 0.9808 & 0.9807 \\
	$1.25\cdot 10^{-3}$  & 0.9911 & 0.9912 \\
	$1.00\cdot 10^{-4}$  & 0.9955 & 0.9955 \\
	$1.00\cdot 10^{-5}$  & 0.9972 & 0.9972 \\
	$1.00\cdot 10^{-6}$  & 0.9981 & 0.9981 \\
	\end{tabular}
	\caption{\label{tab1} Predictions of the analytical theory for the virial potential-energy correlation coefficient $R$ at low densities where the system is in the gas phase (Appendix I). The simulation results are averages over all gas state points at the given temperature (compare \fig{Fig1}).}
\end{table}

At small densities the EXP system is a gas in which individual pair interactions (collisions) dominate the physics. In this limit it is possible to calculate $R$ analytically assuming that the particle collisions are random and uncorrelated. The derivation, which is given in Appendix I, results in

\be
R
\,=\,\frac{A_3}{\sqrt{A_2\,A_4}}\,
\ee
in which (with $\beta\equiv \varepsilon/k_BT$)

\be
A_n
\,=\,\int_{0}^{\infty} v\,\ln^n(1/v)\,e^{-\beta v}dv\,.
\ee
Table \ref{tab1} compares the theory's predictions to numerical results for $R$, which at each temperature have been averaged over the simulated gas-phase state points.

\section{Thermodynamics}\label{IIIa}

\begin{figure}[!htbp]
	\centering
	\includegraphics[width=0.45\textwidth]{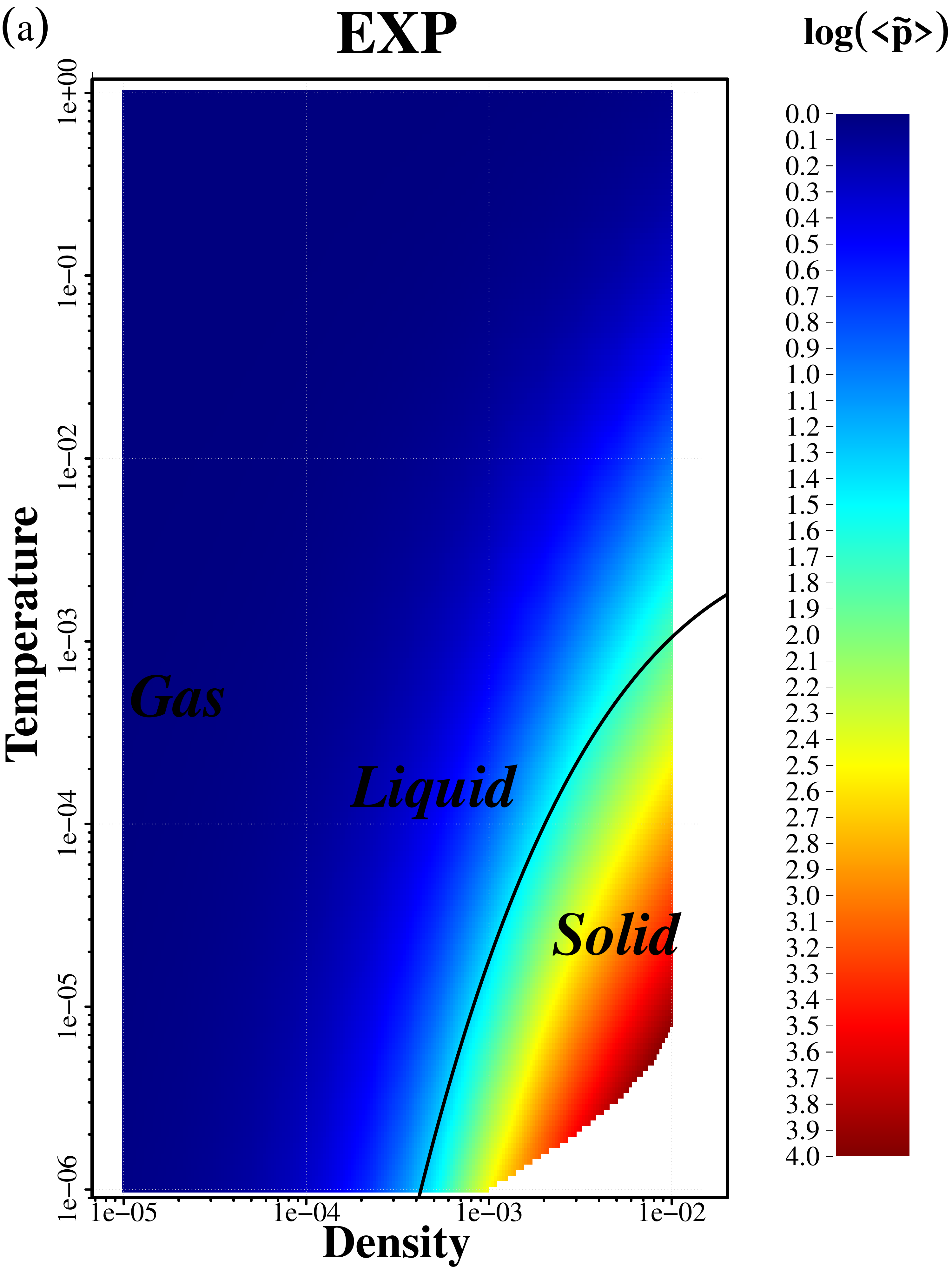}
	\includegraphics[width=0.45\textwidth]{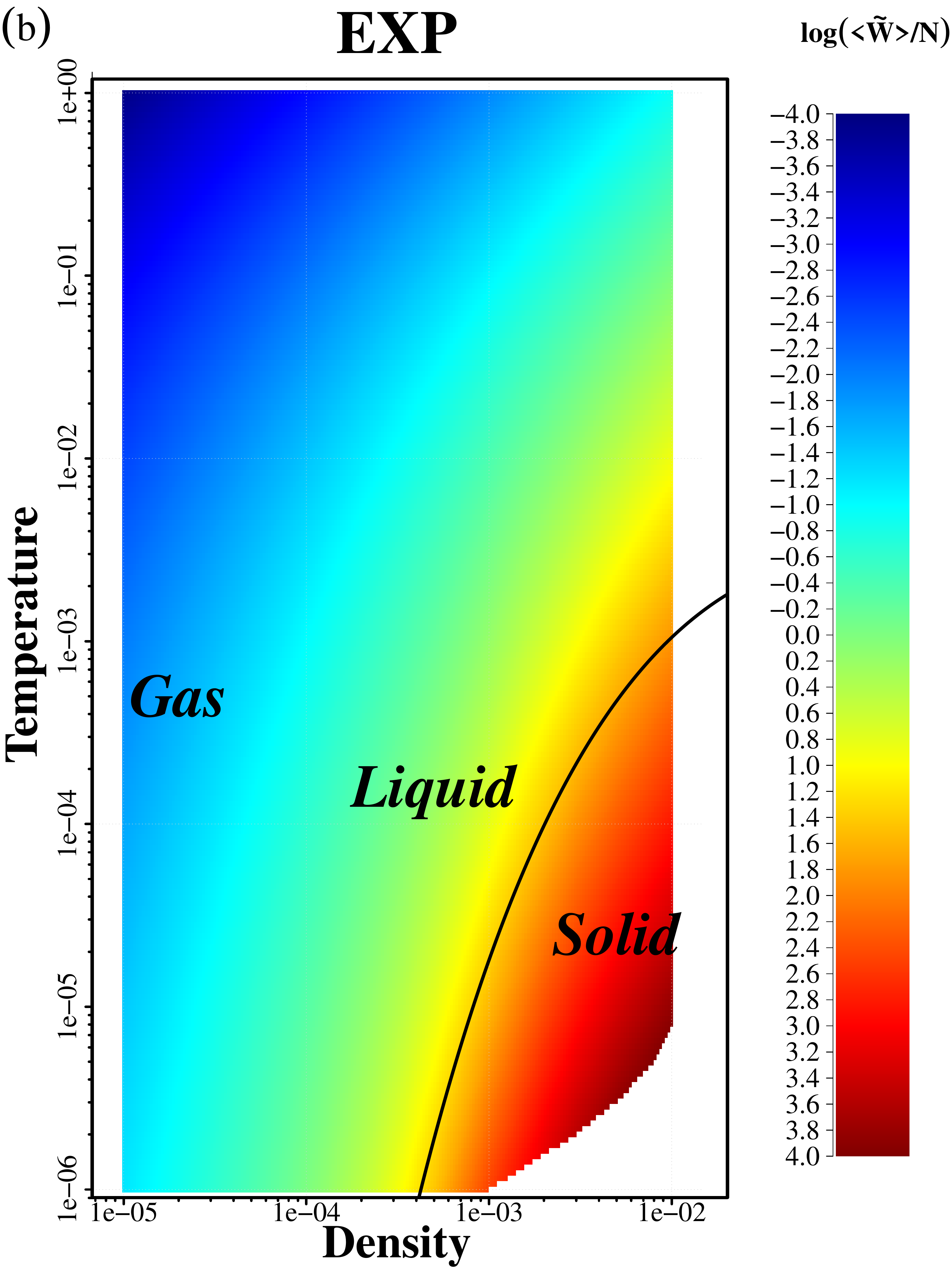}
	\includegraphics[width=0.45\textwidth]{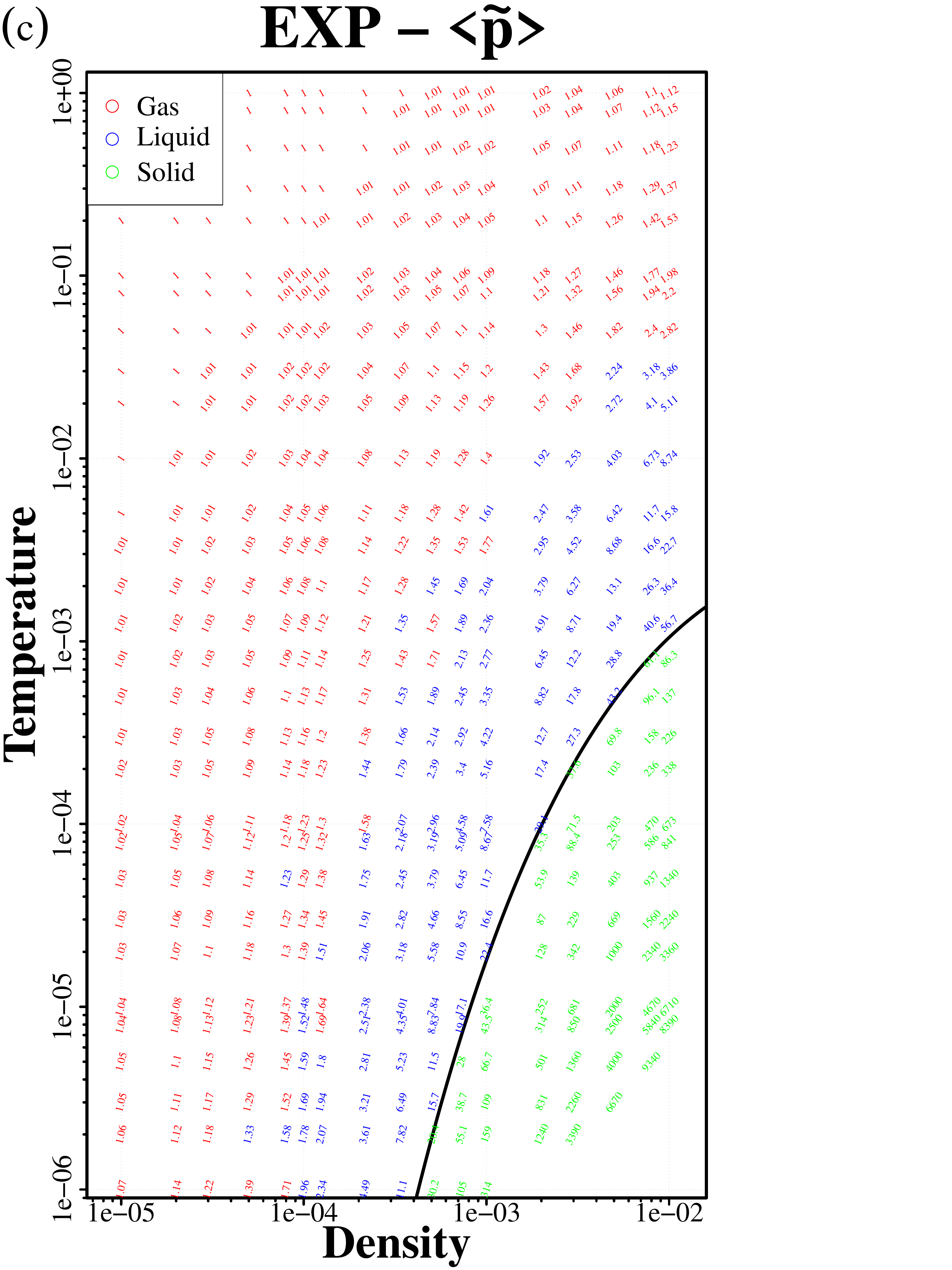}
	\includegraphics[width=0.45\textwidth]{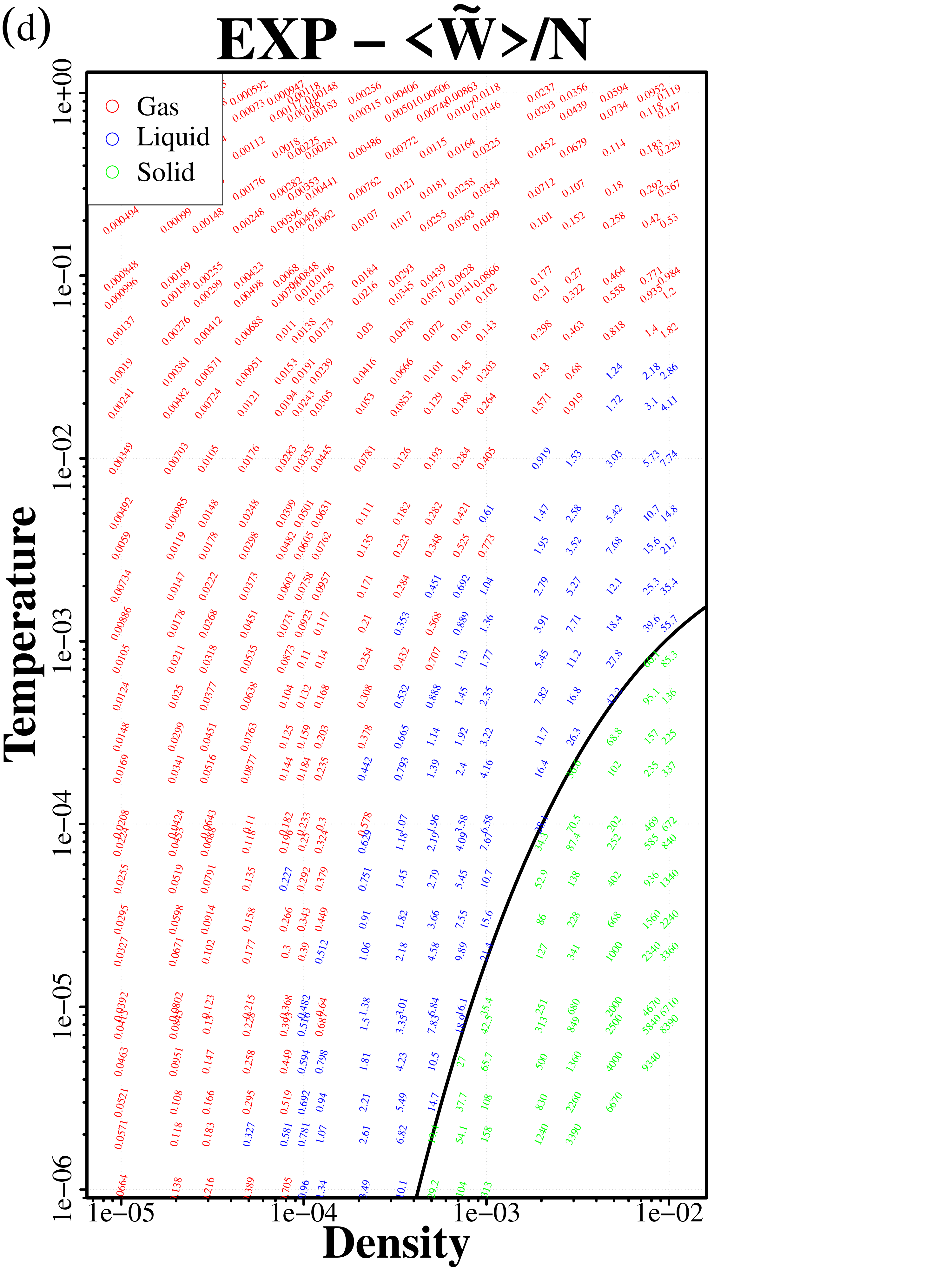}
	\caption{\label{pW_fig} 
	(a) Variation of the logarithm of the average reduced pressure  $\langle\tilde{p}\rangle$.
	(b) Logarithm of the reduced virial per particle $\langle\tilde{W}\rangle/N$.
	(c) and (d) give the numerical values (visible upon magnification) at the densities and temperatures listed in Appendix II. The numerical values are written with a slope marking the direction of the isomorph through the state point in question; red indicates gas, blue liquid, and green solid phase state points.
	}
\end{figure}

An overview of the EXP system's equation of state ($pVT$ relation) is provided in \fig{pW_fig} showing how the average reduced pressure $\langle\tilde{p}\rangle\equiv\langle p\rangle/(\rho k_BT)$ and the average reduced virial per particle $\langle\tilde{W}\rangle/N$ vary throughout the phase diagram. Both quantities are colored after the value of their logarithm. Not surprisingly, the reduced pressure is close to unity in the gas phase ($\tilde{p}=1$ corresponds to the ideal gas equation), but it grows and becomes much larger as the liquid and solid phases are approached. Comparing \fig{pW_fig}(a) and (b) reveals that the virial per particle in the gas phase is much lower than the pressure, whereas in the solid and liquid phases the pressure is dominated by the virial. For reference, (c) and (d) report the numerical values of average pressure and virial.

\begin{figure}[!htbp]
	\centering
	\includegraphics[width=0.45\textwidth]{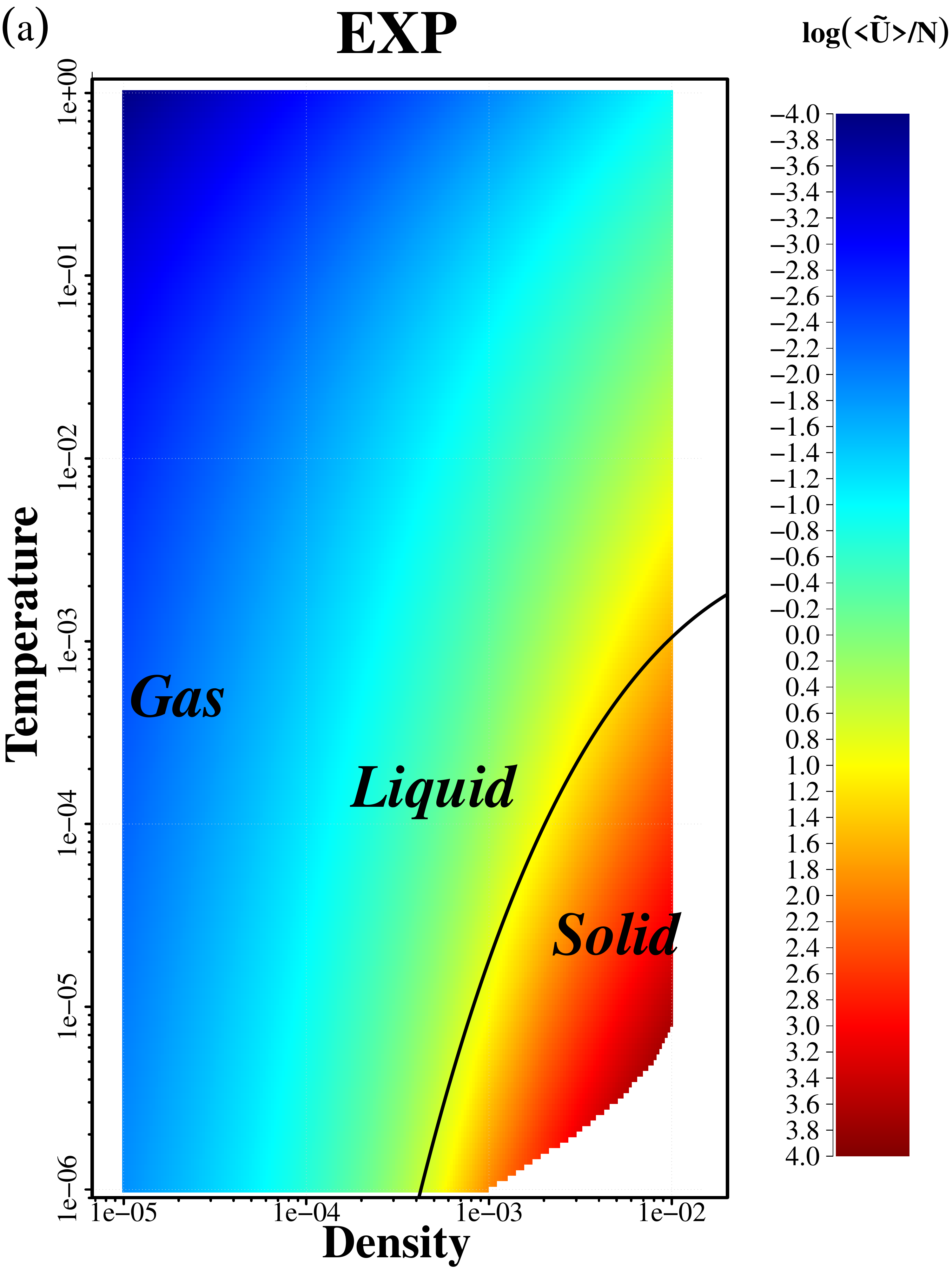}
	\includegraphics[width=0.45\textwidth]{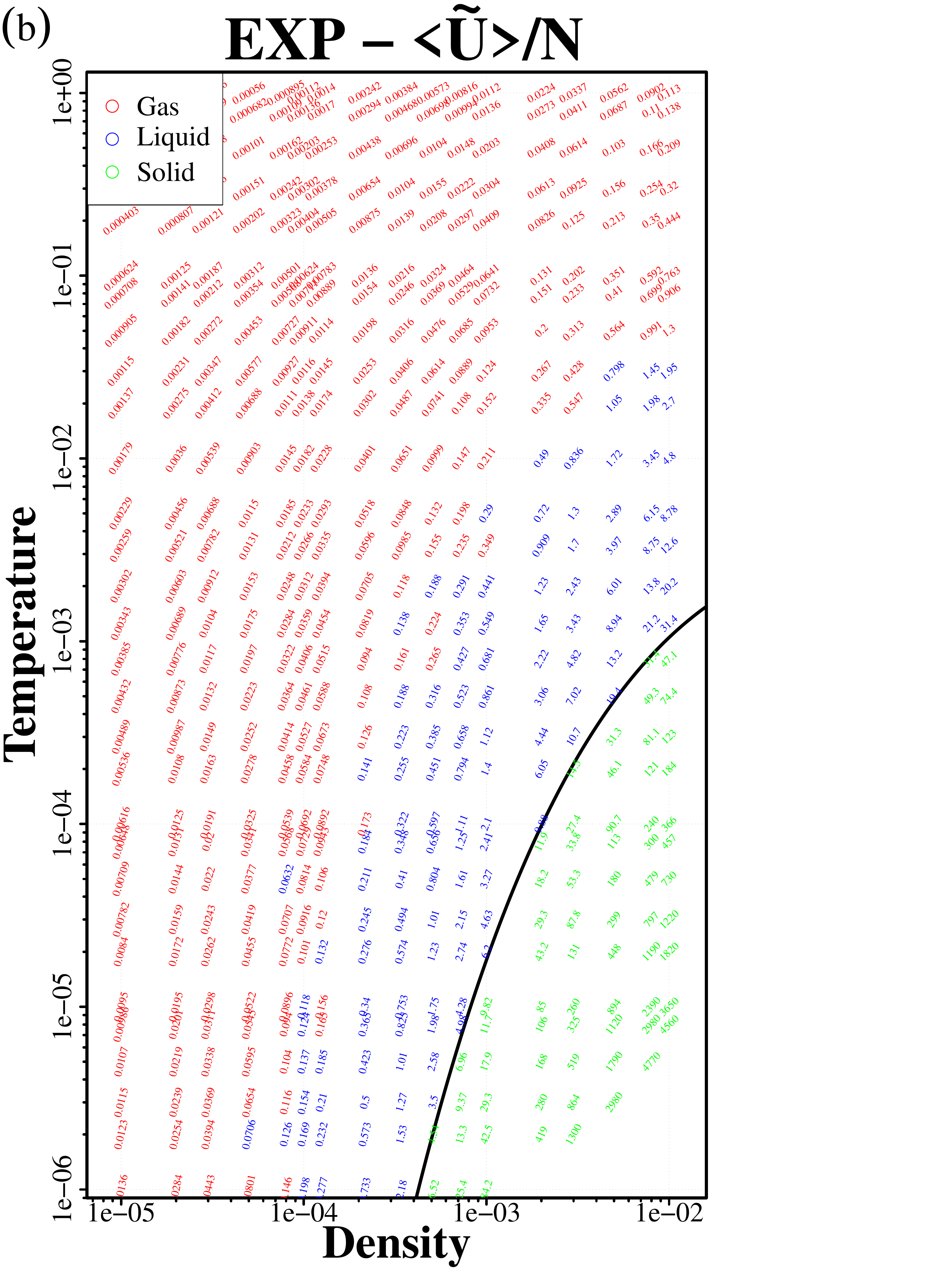}
	\caption{\label{U_fig}
	(a) Variation of the logarithm of the reduced potential energy throughout the phase diagram.
	(b) gives the numerical values of $\tilde{U}$ at different state points (visible upon magnification) at the densities and temperatures listed in Appendix II. At each state point the numerical value is written with a slope marking the direction of the isomorph through the state point in question; red indicates gas, blue liquid, and green solid phase state points. }
\end{figure}

The variation of the reduced average potential energy per particle is shown in \fig{U_fig}(a). The gas phase is characterized by a much lower potential energy than the kinetic energy, implying $\langle\tilde{U}\rangle/N\ll 1$. In the solid phase the opposite behavior is seen; here the potential energy dominates.

\section{Structure, dynamics, and specific heat along isotherms}\label{IV}

This section investigates the EXP system's properties along selected isotherms, the next section does the same along isochores. Both sections cover temperatures between $10^{-6}$ and $1$ and densities between $10^{-5}$ and $10^{-2}$, with a focus on the gas and liquid phases (compare \fig{Fig1}).

\begin{figure}[!htbp]
	\centering
	\includegraphics[width=0.45\textwidth]{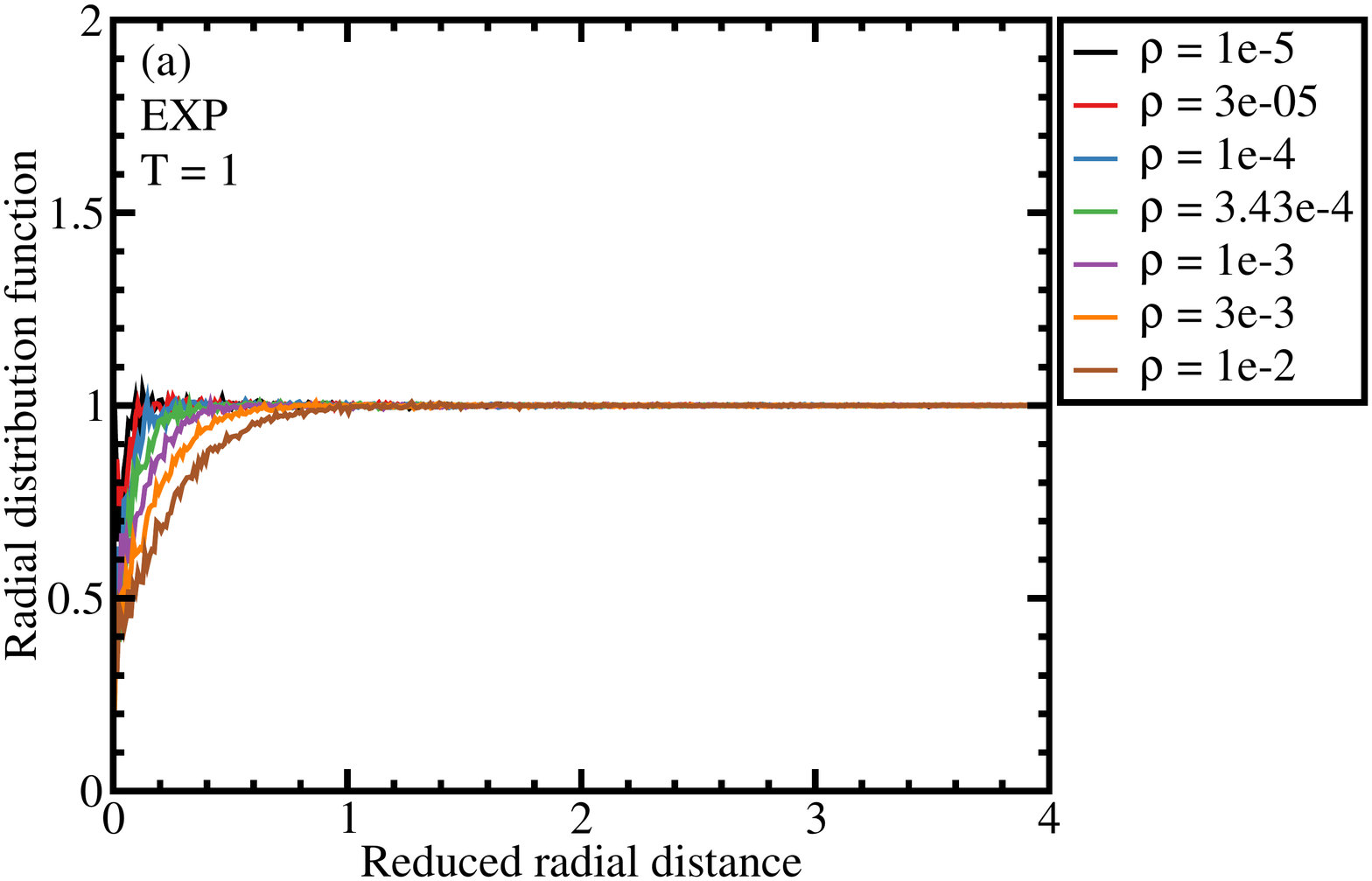}
	\includegraphics[width=0.45\textwidth]{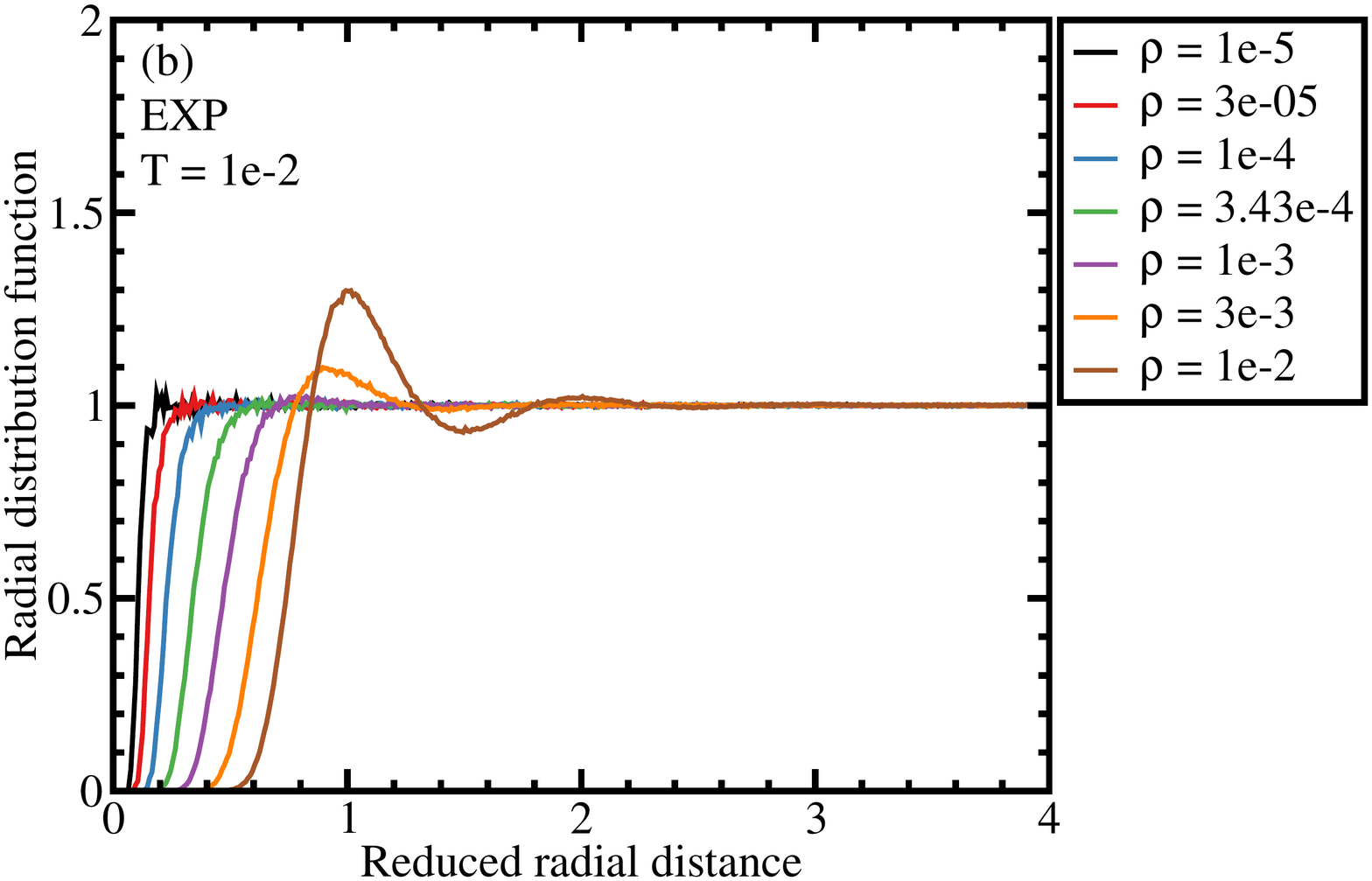}
	\includegraphics[width=0.45\textwidth]{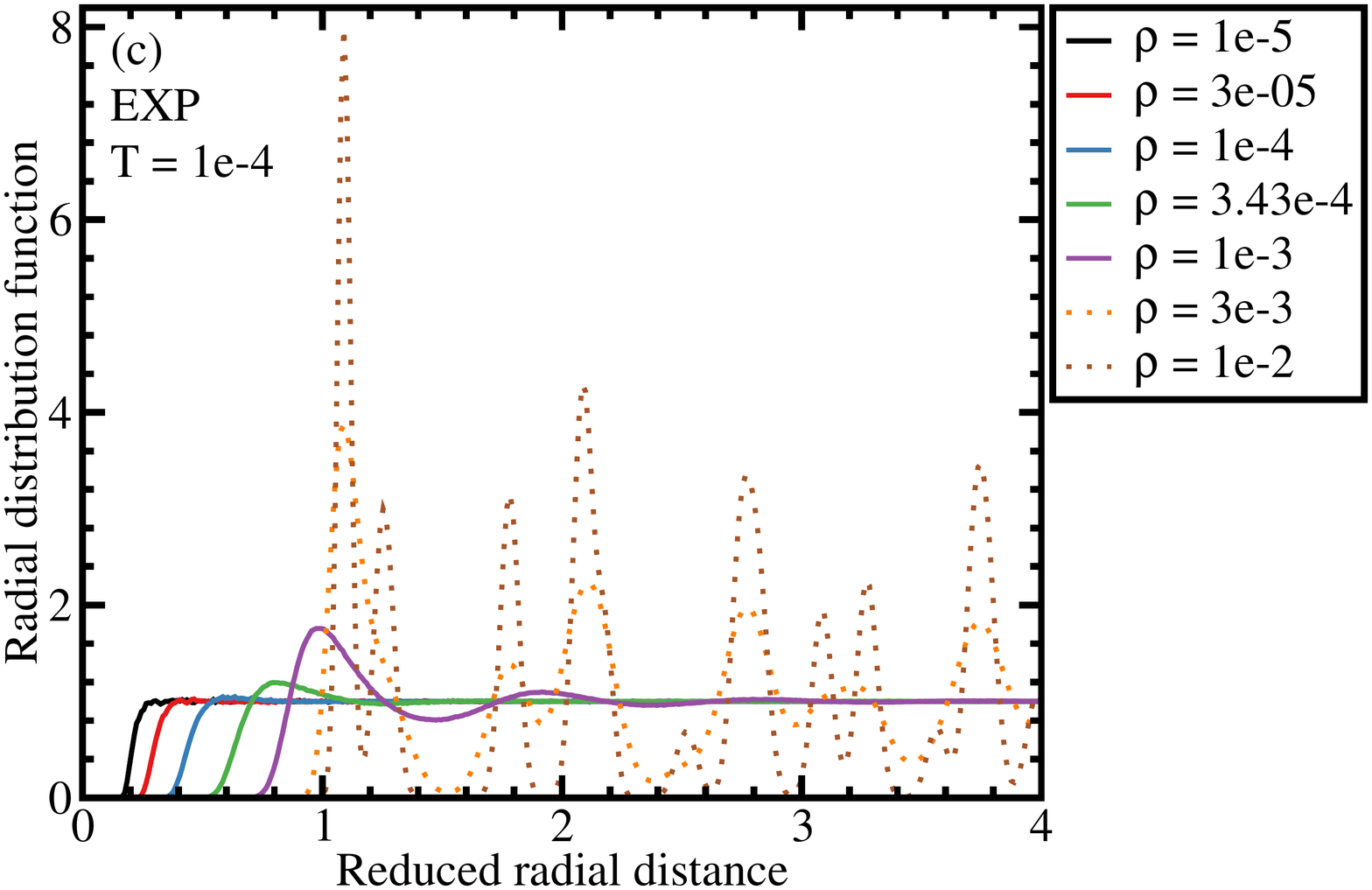}
	\includegraphics[width=0.45\textwidth]{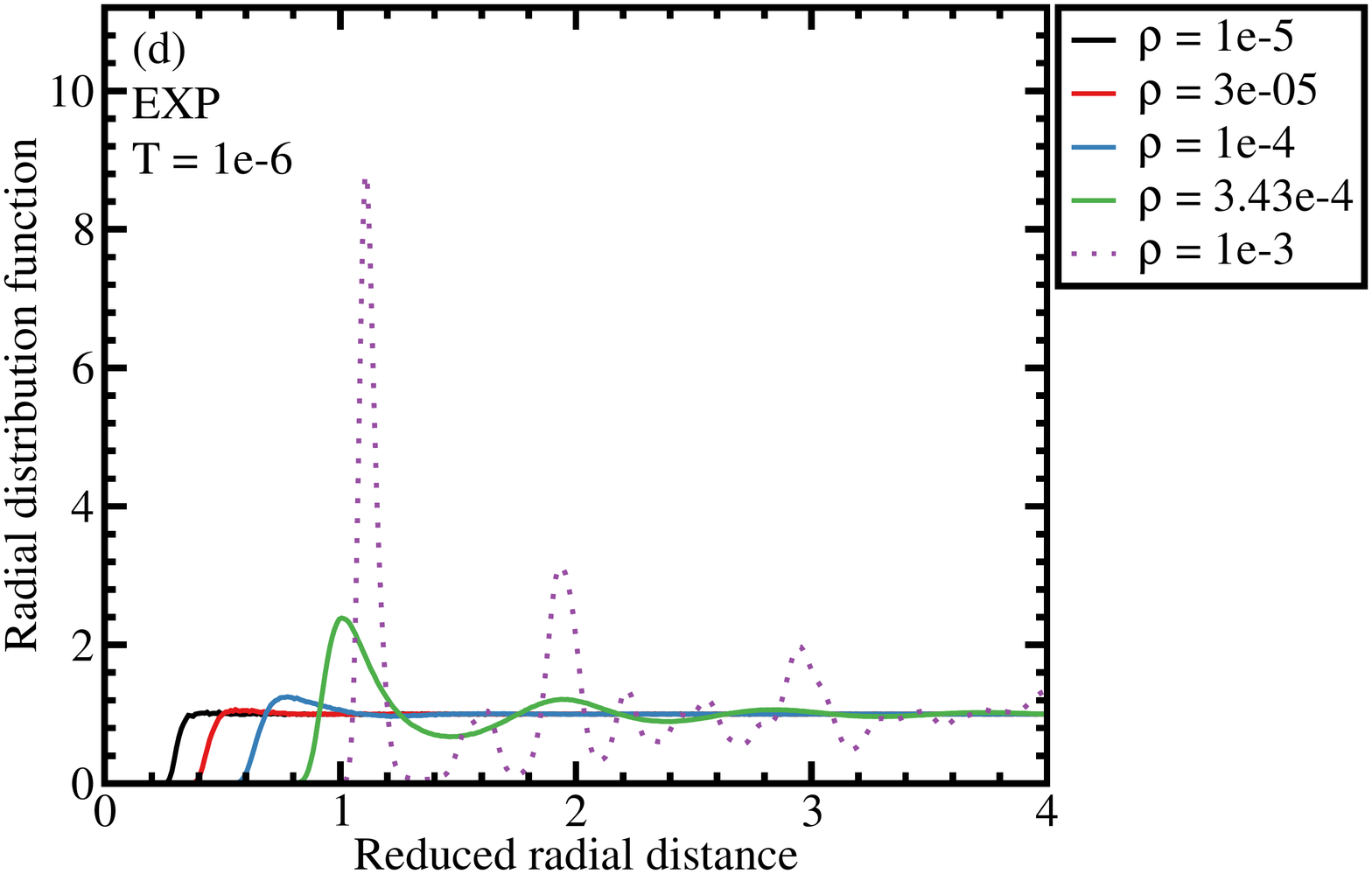}
	\caption{\label{fi:isoT-rdf} Structure along isotherms probed by the radial distribution function (RDF) as a function of the reduced pair distance $\tilde{r}\equiv\rho^{1/3}r$ at the following temperatures: (a) $T = 1$, (b) $T = 10^{-2}$, (c) $T= 10^{-4}$, (d) $T = 10^{-6}$. The RDFs of state points in the gas and liquid phases (\fig{Fig1}) are indicated by full lines, solid-phase RDFs by dotted lines. 
		(a) corresponds to the average kinetic energy per particle comparable to the pair potential energy at zero separation. In this case the system is gas-like over the entire density range investigated. 
		(b) Liquid-like structure is observed at the highest densities. 
		(c) and (d) With increasing density the system transforms from gas to liquid to solid behavior.}
\end{figure}

\Fig{fi:isoT-rdf} shows how the radial distribution function (RDF) $g(r)$ develops with density at four temperatures: $T=1$, $T=10^{-2}$, $T=10^{-4}$, and $T=10^{-6}$. Recall that the RDF gives the probability to find two particles the distance $r$ from each other relative to that of an ideal gas at the same density. When comparing $g(r)$ at different state points it is convenient to use reduced units, $\tr\equiv\rho^{1/3}r$. 

At the highest temperature $T=1$ there is little structure; here the system is a gas at all densities investigated. At close distances the RDF falls below unity, reflecting the interparticle repulsion. Because of the reduced units used, at low densities this happens for $\tilde{r}\ll 1$. (b) shows $T=10^{-2}$ data; some structure appears at the highest densities. (c) shows the data for  $T=10^{-4}$. The range of densities studied here comprise a few solid state points revealed as spikes in the RDFs that are present also at large distances (dashed lines). (d) gives RDFs for $T=10^{-6}$ at which a similar pattern appears. For all four temperatures the low-density state points have little structure; they are all in the gas phase.

	\begin{figure}[!htbp]
		\centering
		\includegraphics[width=0.45\textwidth]{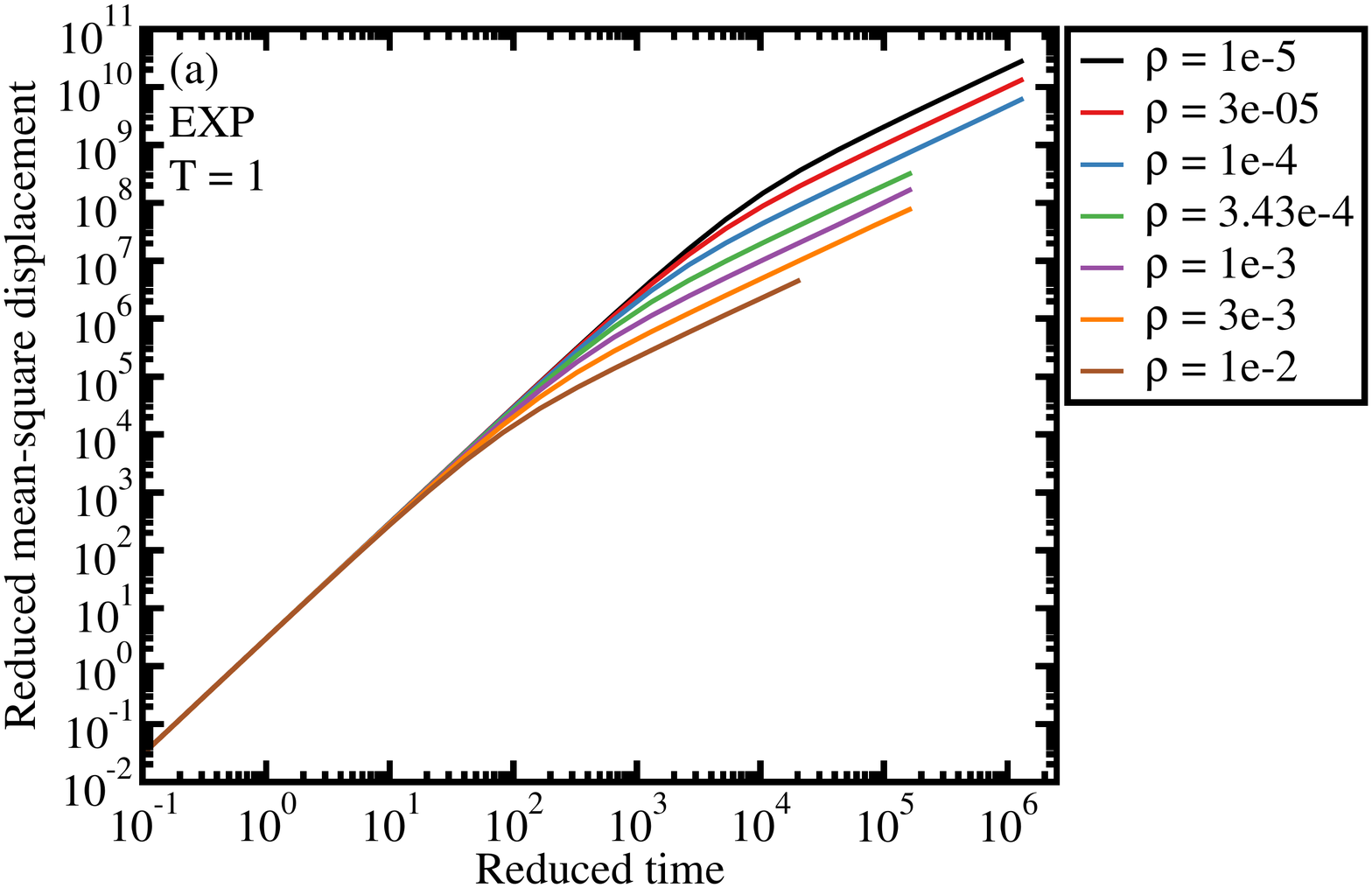}
		\includegraphics[width=0.45\textwidth]{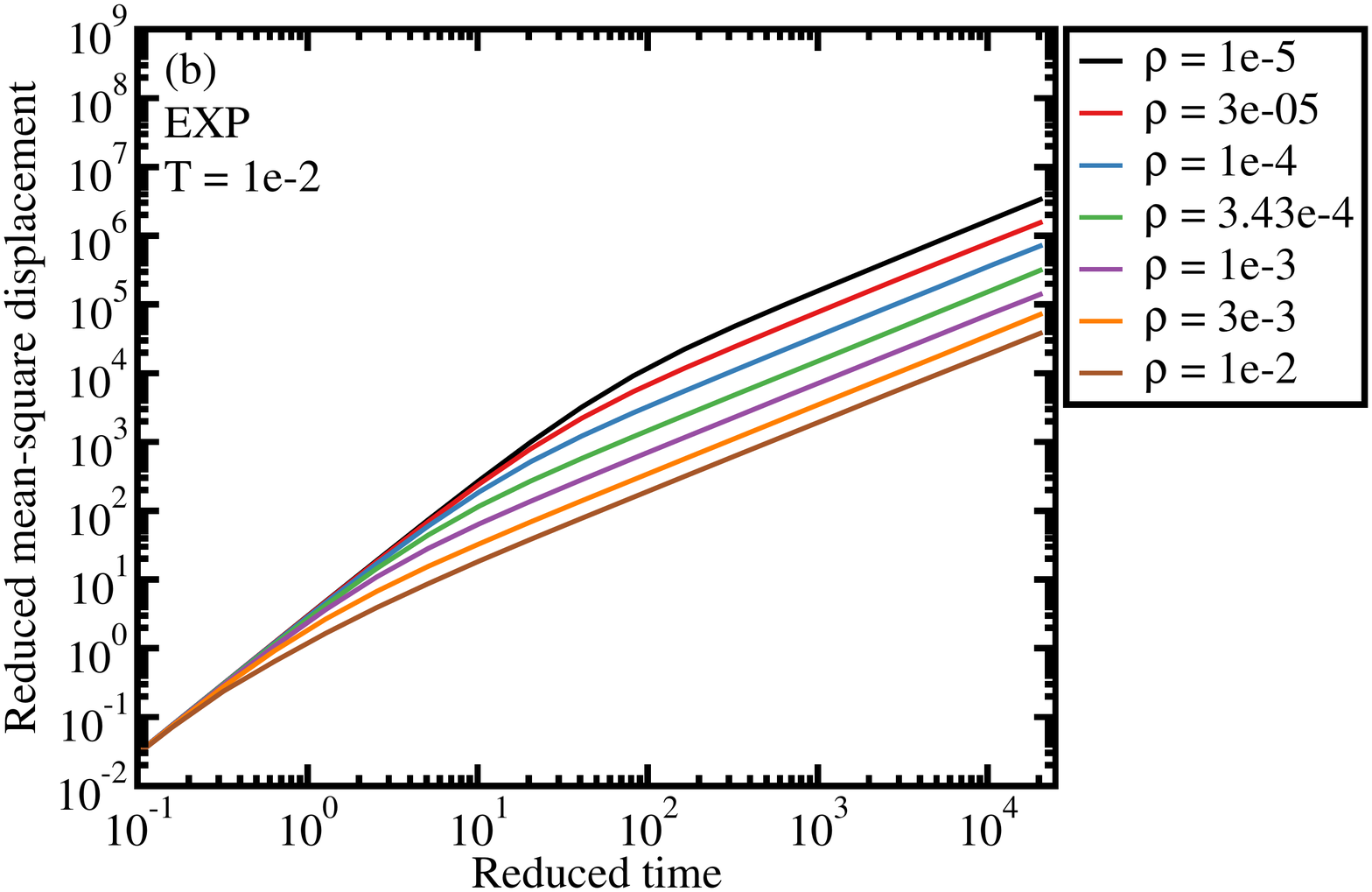}
		\includegraphics[width=0.45\textwidth]{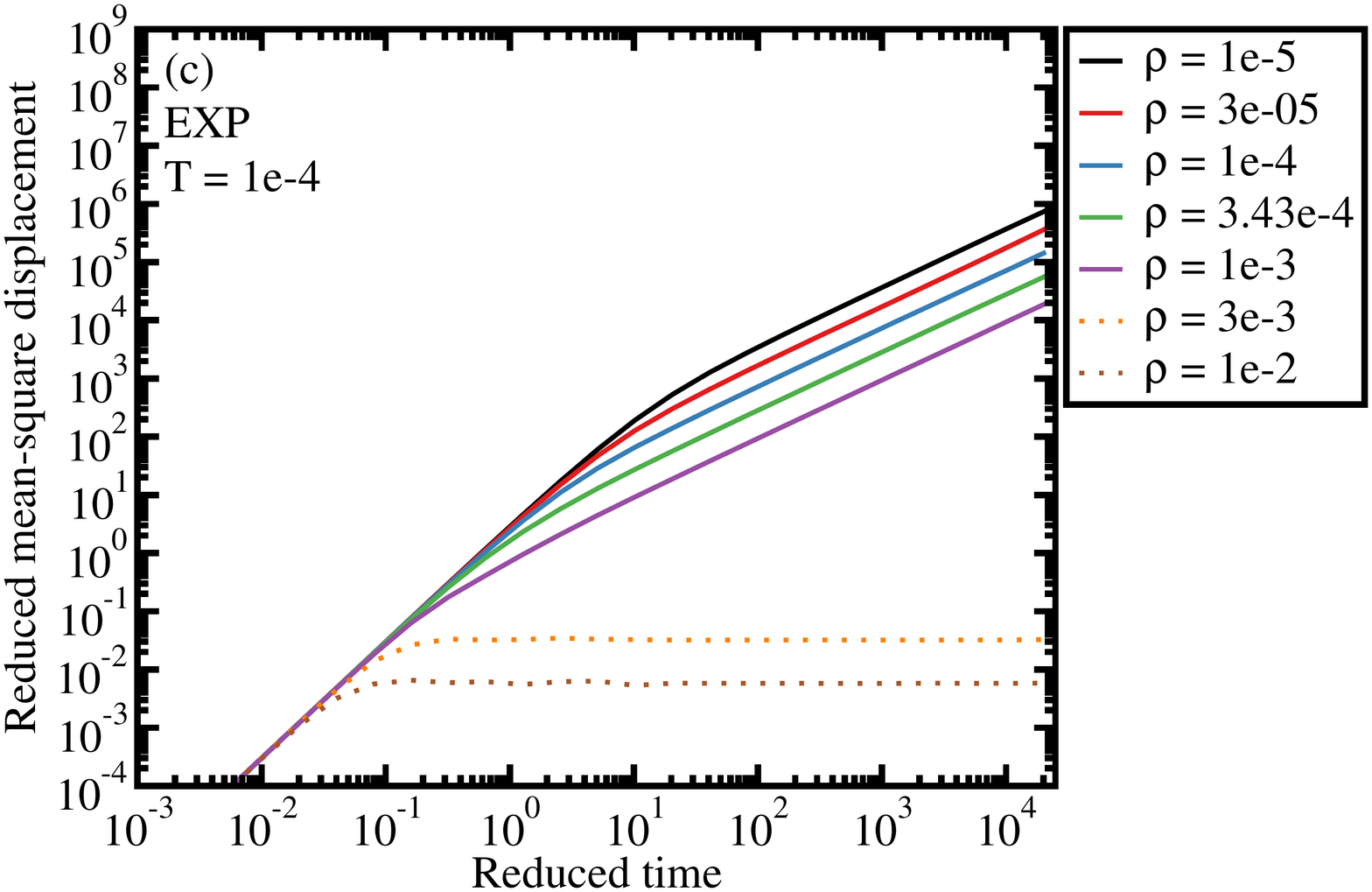}
		\includegraphics[width=0.45\textwidth]{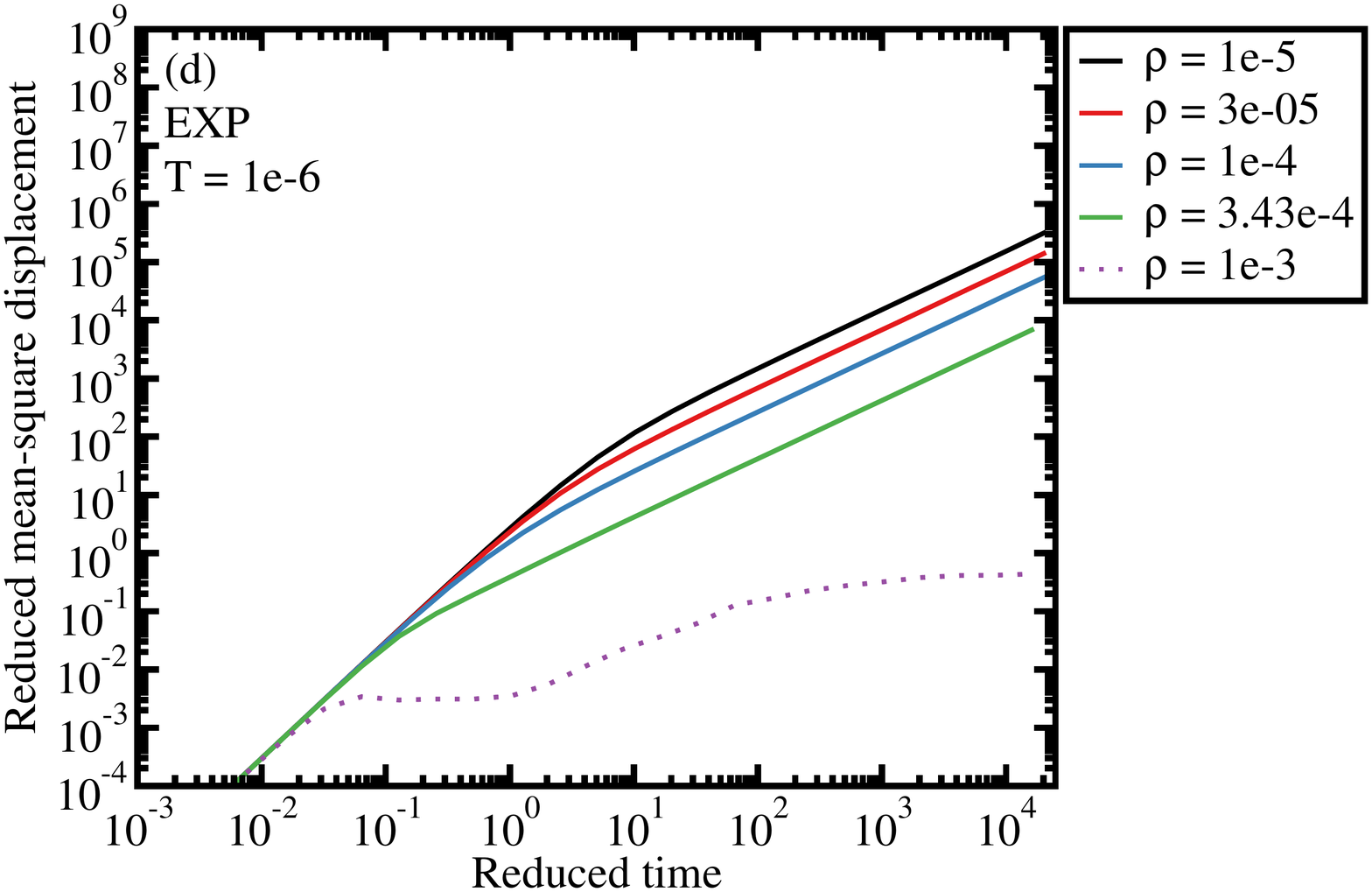}
		\caption{\label{fi:isoT-msd}Reduced-unit mean-square displacement (MSD) for selected state points along the four isotherms $T = 1$, $T = 1\cdot 10^{-2}$, $T = 1\cdot 10^{-4}$, $T = 1\cdot 10^{-6}$. Gas and liquid phase state points are indicated by full lines, solid state points by dotted lines. At short times the MSD in all cases follows the ballistic prediction $3\,\tilde t^2$ (see the text), at long times it follows the diffusion equation prediction $\propto \tilde t$. The solid phase does not reach the long-time diffusive limit.}
	\end{figure}

\Fig{fi:isoT-msd} shows the mean-square displacement (MSD) $\langle\Delta r^2(t)\rangle$ as a function of time evaluated along the same four isotherms. The MSD is converted into reduced units by multiplying by $\rho^{2/3}$ while time is multiplied by $\rho^{1/3}\sqrt{k_BT/m}$ (the inverse of the time for a free particle of kinetic energy $k_BT$ to move a nearest-neighbor distance). (a) gives the $T=1$ results for the same range of densities as in \fig{fi:isoT-rdf}. At short times corresponding to ballistic motion the reduced MSD equals $3\,\tilde t^2$ since 
$\langle\Delta \tilde r^2(t)\rangle=\rho^{2/3}\langle {\bf v}^2\rangle t^2=\rho^{2/3}3(k_BT/m)t^2=3{\tilde {t}}^2$. At long times the MSD varies in proportion to $\tilde t$, which is the well-known diffusive motion. The transition to diffusive motion takes place later, the lower the density is. This is because for gas-like states the mean free path $l$ is much larger than the average nearest-neighbor distance \cite{chapman,kau66} (see below). At lower temperature ((b)) the transition moves closer to $\tt\sim 1$ as density increases. In (c) and (d) reporting results for the two lowest temperatures we observe at high densities a solid phase MSD (dashed lines, not equilibrated). 

	\begin{figure}[!htbp]
	\centering
	\includegraphics[width=0.45\textwidth]{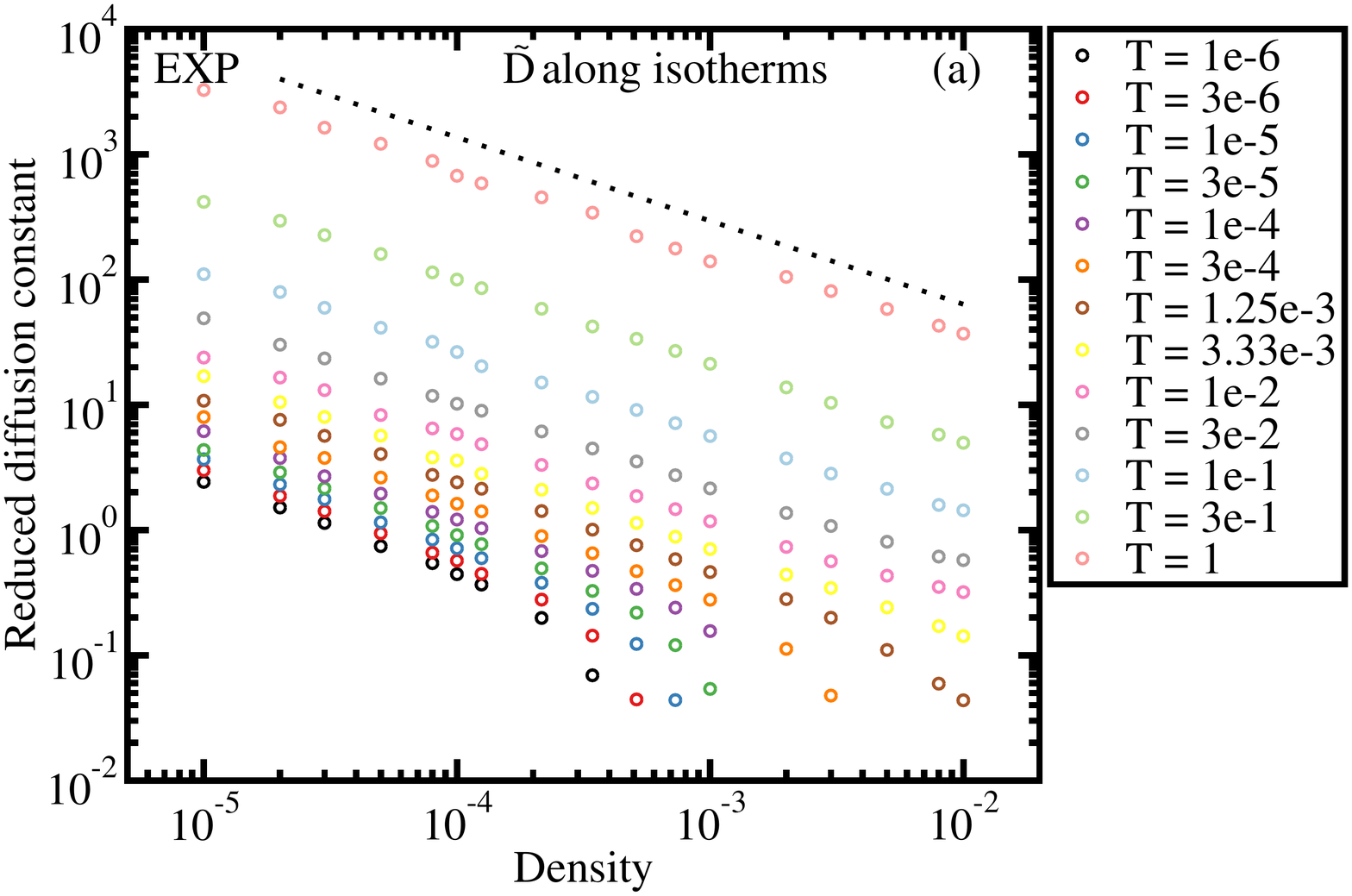}
	\includegraphics[width=0.45\textwidth]{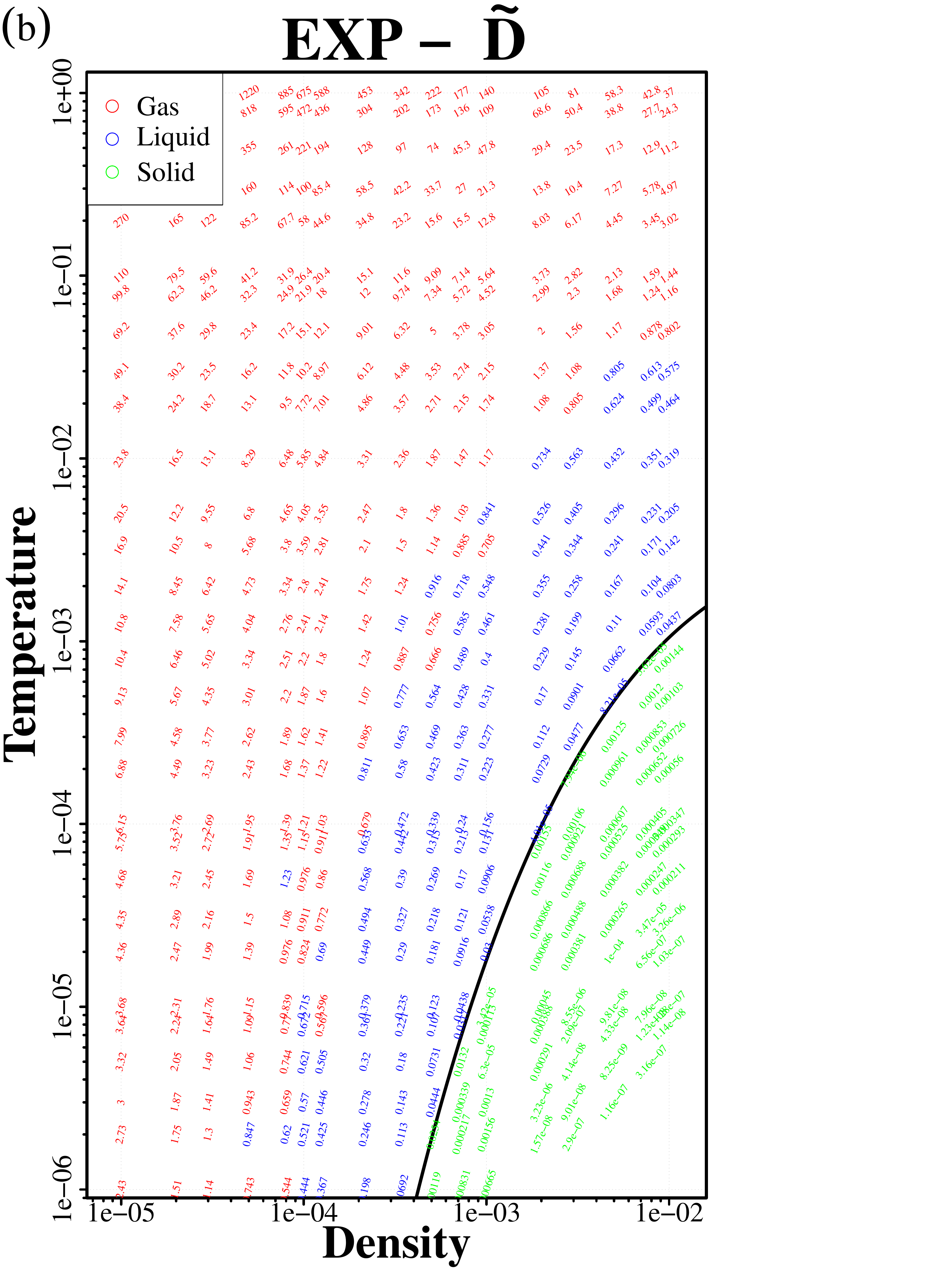}
	\caption{\label{fi:isoT-D} Reduced diffusion constants determined from the long-time limit of the MSD.
	(a) shows results for isotherms as a function of the density. At all temperatures the reduced diffusion constant decreases as density increases; at high temperatures one finds $\tilde{D}\propto\rho^{-2/3}$ as predicted by kinetic theory (\eq{enskog}, dashed line).
	(b) gives the numerical values of the reduced diffusion constants (visible upon magnification) at the same state points as those of \fig{fi:phase-diagram} (Appendix II). Red are gas, blue liquid, and green solid phase state points. At each state point the value of the reduced diffusion coefficient is written with a slope marking the direction of the isomorph through the state point in question.}
	\end{figure}

\Fig{fi:isoT-D}(a) shows the reduced diffusion constant $\tilde{D}$ derived from long-time MSD data via $\langle\Delta r^2(t)\rangle=6Dt$ along several isotherms in the gas and liquid phases. At fixed temperature $\tilde{D}$ decreases when density increases; the mean-free path is reduced and approaches the average interparticle distance as the gas phase transforms smoothly into the liquid phase. More accurately, as we now proceed to show, one has in the gas phase $\tilde D\propto\rho^{-2/3}$ at fixed temperature (dashed line in \fig{fi:isoT-D}(a)).

According to kinetic theory \cite{chapman,kau66}, the diffusion constant in the gas phase is proportional to $\,lv\,$ in which $l$ is the mean free path and $v$ the thermal velocity. Since $v$ is basically $l_0/t_0$, this implies $\tilde{D}\equiv D/(l_0^2/t_0)=l/l_0\propto l\rho^{1/3}$. Gas-phase kinetic theory moreover predicts that $l$ is given by $\rho l r_0^2\sim 1$ where $r_0$ is the effective particle radius that may be estimated from $v_{\rm EXP}(-r_0) = k_BT$, implying $ r_0 = \ln(1/T)=- \ln T$ in the EXP unit system. Thus one expects the reduced diffusion constant in the gas phase to be given by $\tilde{D}\,\propto\,l\rho^{1/3}\,\propto\,\rho^{-2/3}/r_0^2\,=\,\rho^{-2/3}/\ln^2(T)$. The Enskog kinetic theory determines the constant of proportionality \cite{chapman,kau66}, resulting in

\be\label{enskog}
\tilde{D}
\,=\,\frac{3}{8\sqrt{\pi}}\,\frac{\rho^{-2/3}}{\ln^2(T)}
\,=\,0.212\,\frac{\rho^{-2/3}}{\ln^2(T)}\,.
\ee
This expression is validated below in \fig{fi:isor-D}. Note that the effective hard-sphere approximation is expected work best at low densities and low temperatures, which is consistent with the findings of \fig{fi:isoT-D}(a).

	\begin{figure}[!htbp]
		\centering
		\includegraphics[width=0.45\textwidth]{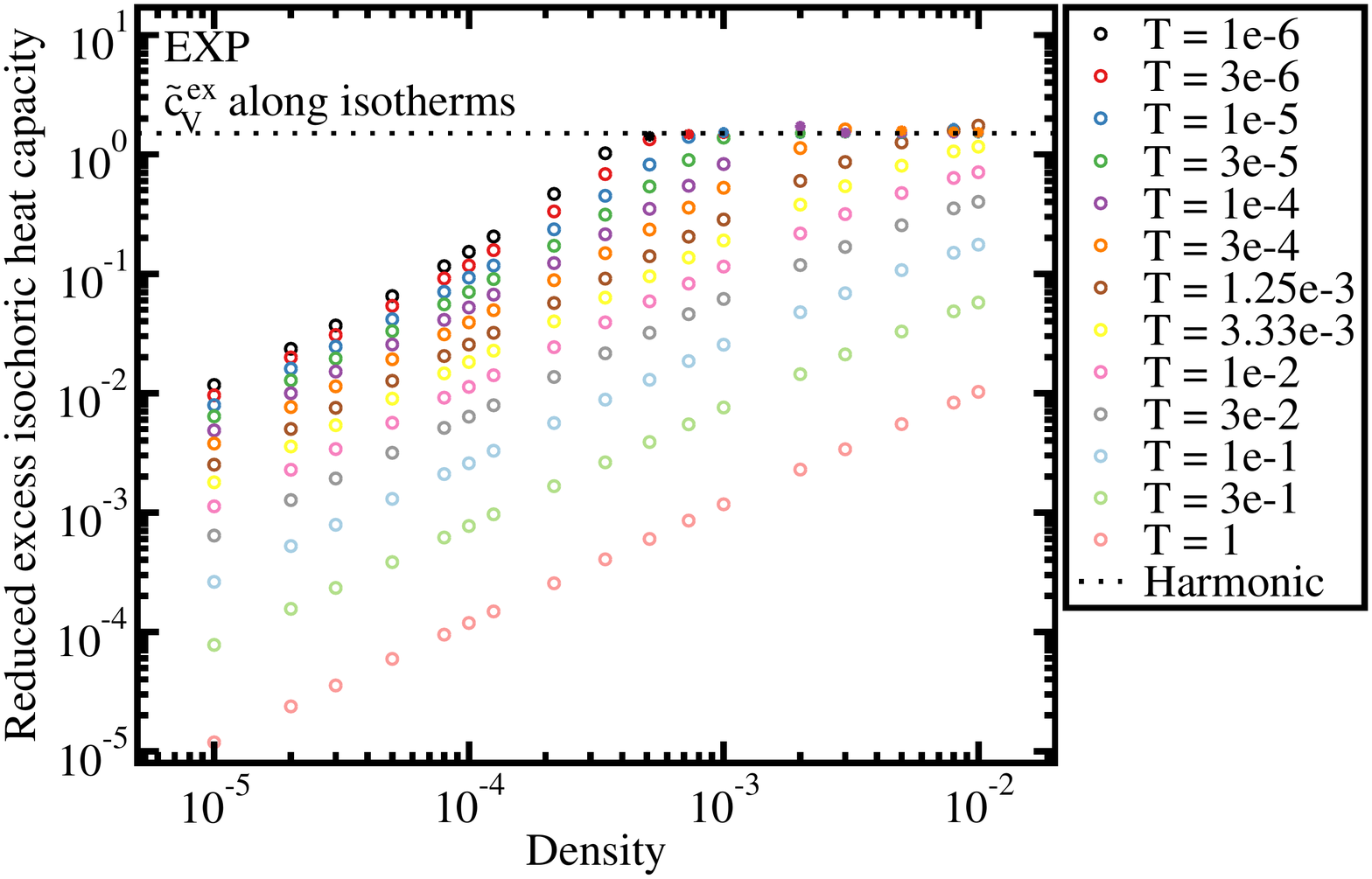}
		\caption{\label{fi:isoT-C}Reduced isochoric excess heat capacity per particle along  isotherms. Gas and liquid state points are given as open symbols, stars represent crystal state points. The dashed line is the prediction for a harmonic crystal, $\tcVex=3/2$.}
	\end{figure}

\Fig{fi:isoT-C} gives the reduced excess isochoric specific heat per particle $\tcVex$, i.e., $c_V/k_B$ subtracted the $3/2$ ideal-gas per-particle contribution. This is calculated from the system's potential-energy fluctuations in $NVT$ simulations via Einstein's canonical-ensemble expression $\CVex=\langle(\Delta U)^2\rangle/k_BT^2$  \cite{tildesley,IV}. The lowest temperatures have the largest $\tcVex$, reflecting stronger interactions than at higher temperatures, which are more gas like. There is a transition to a virtually constant $\tcVex\cong 3/2$ at high densities at which the system is in the crystalline state.

\section{Structure, dynamics, and specific heat along isochores}\label{V}

Next we study how the above quantities vary along the lines of constant volume, reporting results for the densities $10^{-5}$, $10^{-4}$, $10^{-3}$, and $10^{-2}$. 

	\begin{figure}[!htbp]
		\centering
		\includegraphics[width=0.45\textwidth]{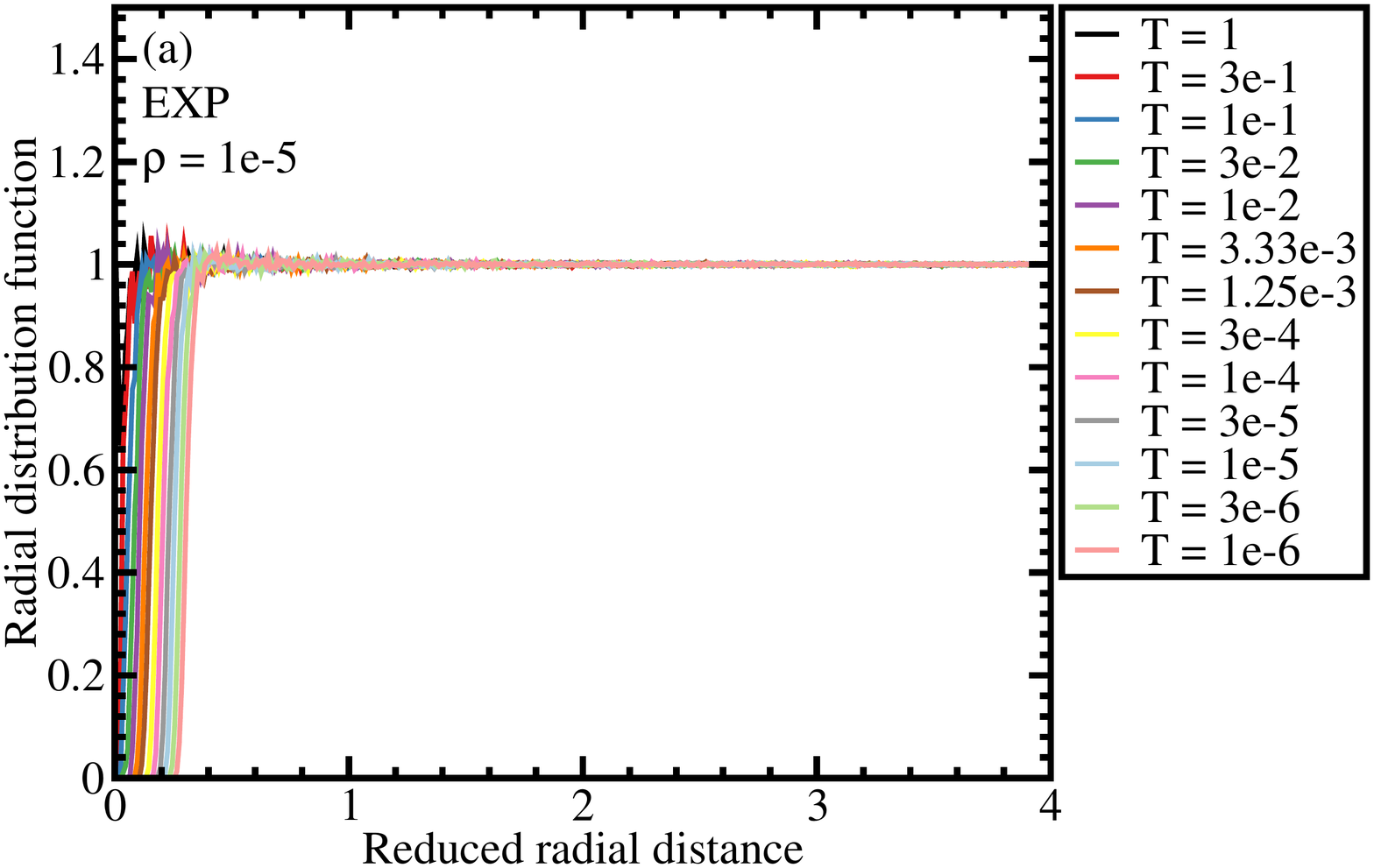}
		\includegraphics[width=0.45\textwidth]{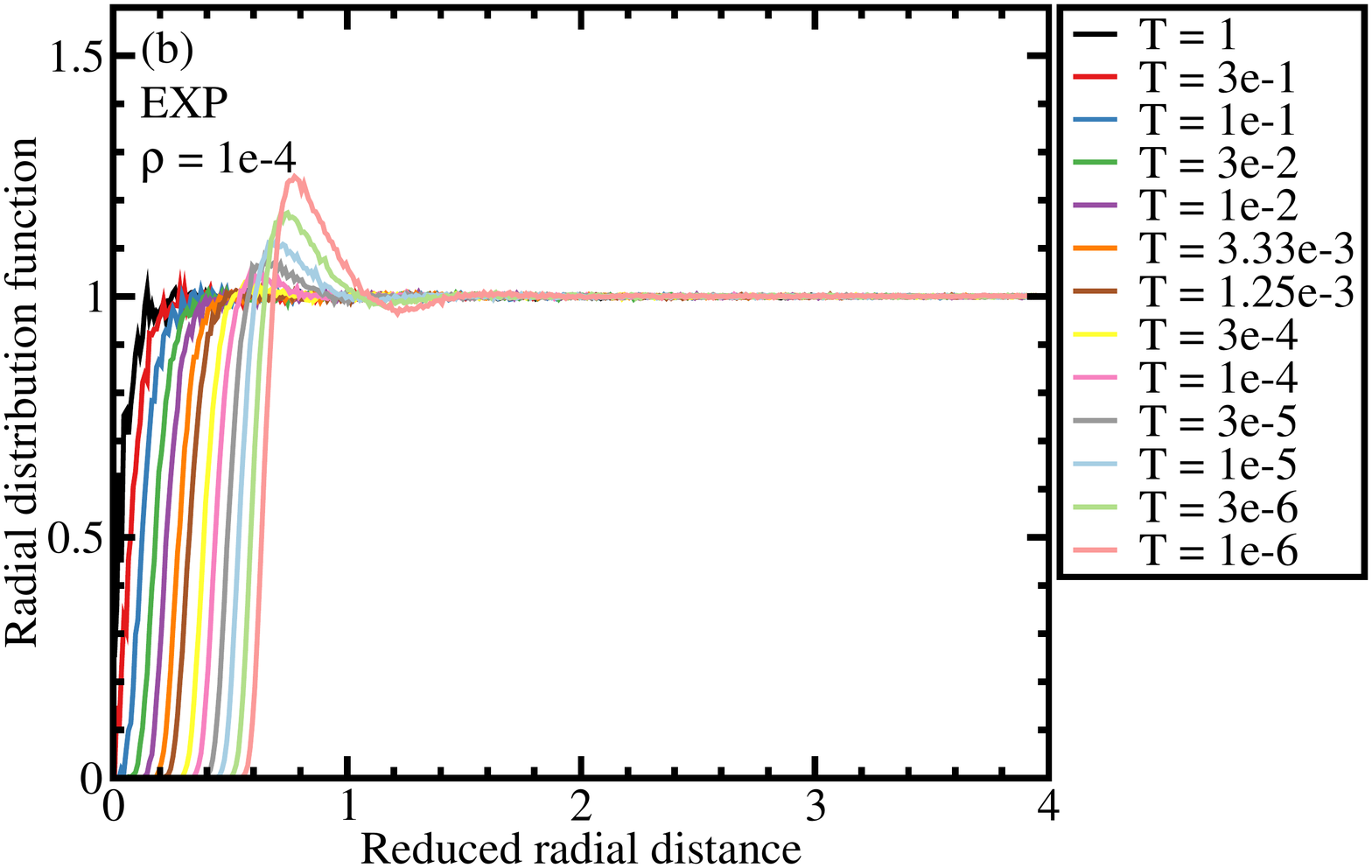}
		\includegraphics[width=0.45\textwidth]{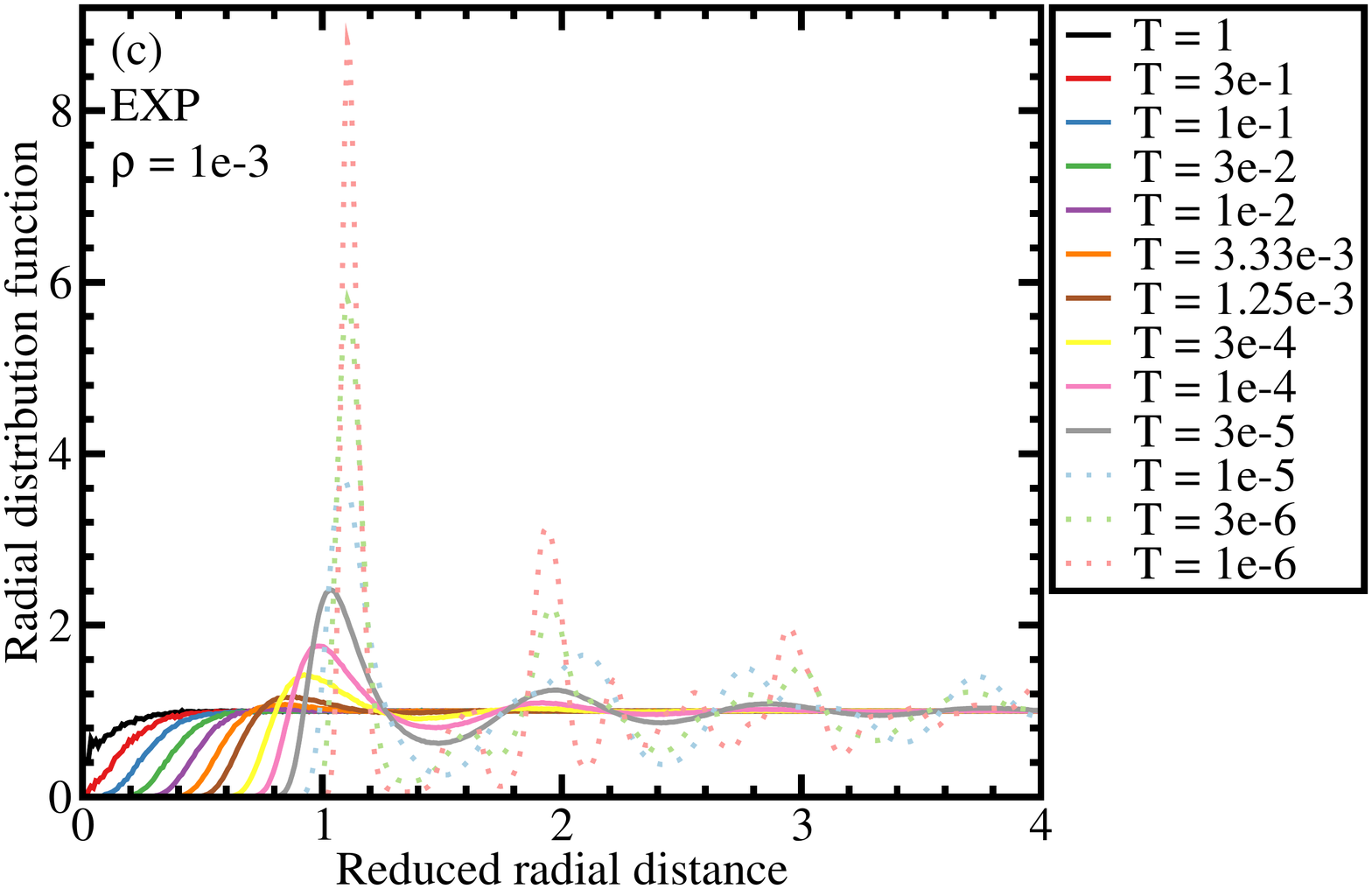}
		\includegraphics[width=0.45\textwidth]{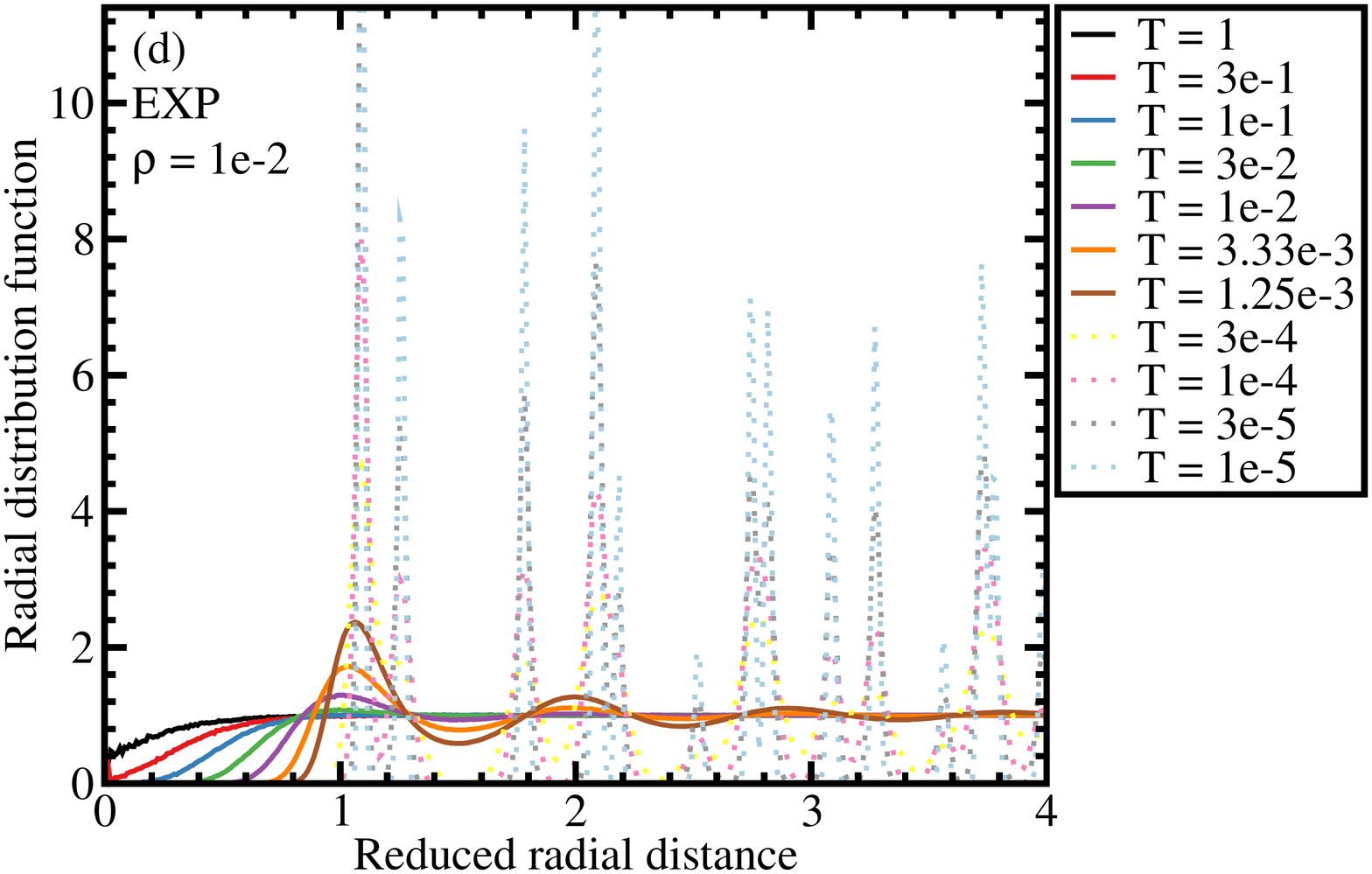}
		\caption{\label{fi:isor-rdf}Radial distribution functions along the following isochores: (a) ${\rho} = 10^{-5}$, (b) ${\rho} = 10^{-4}$, (c)  ${\rho} = 10^{-3}$, (d) ${\rho} = 10^{-2}$. State points in the gas and liquid phases (compare \fig{Fig1}) are indicated by full lines, solid-phase state points by dotted lines. At the lowest densities little structure is present and the system is a gas even at the lowest temperatures studied. At higher densities liquid-like structure appears at the lowest temperatures, and for ${\rho} = 10^{-3}$ and ${\rho} = 10^{-2}$ crystal structure is observed at the lowest temperatures.}
	\end{figure}

\Fig{fi:isor-rdf} shows the reduced RDFs along these four isochores. At the lowest density (a) there is little structure. Here the system is a gas at all temperatures, compare \fig{Fig1}. As density is increased, structure begins to appear at the lowest temperatures, and for the two highest densities we recognize the crystalline phase (dashed lines).

	\begin{figure}[!htbp]
	\centering
	\includegraphics[width=0.45\textwidth]{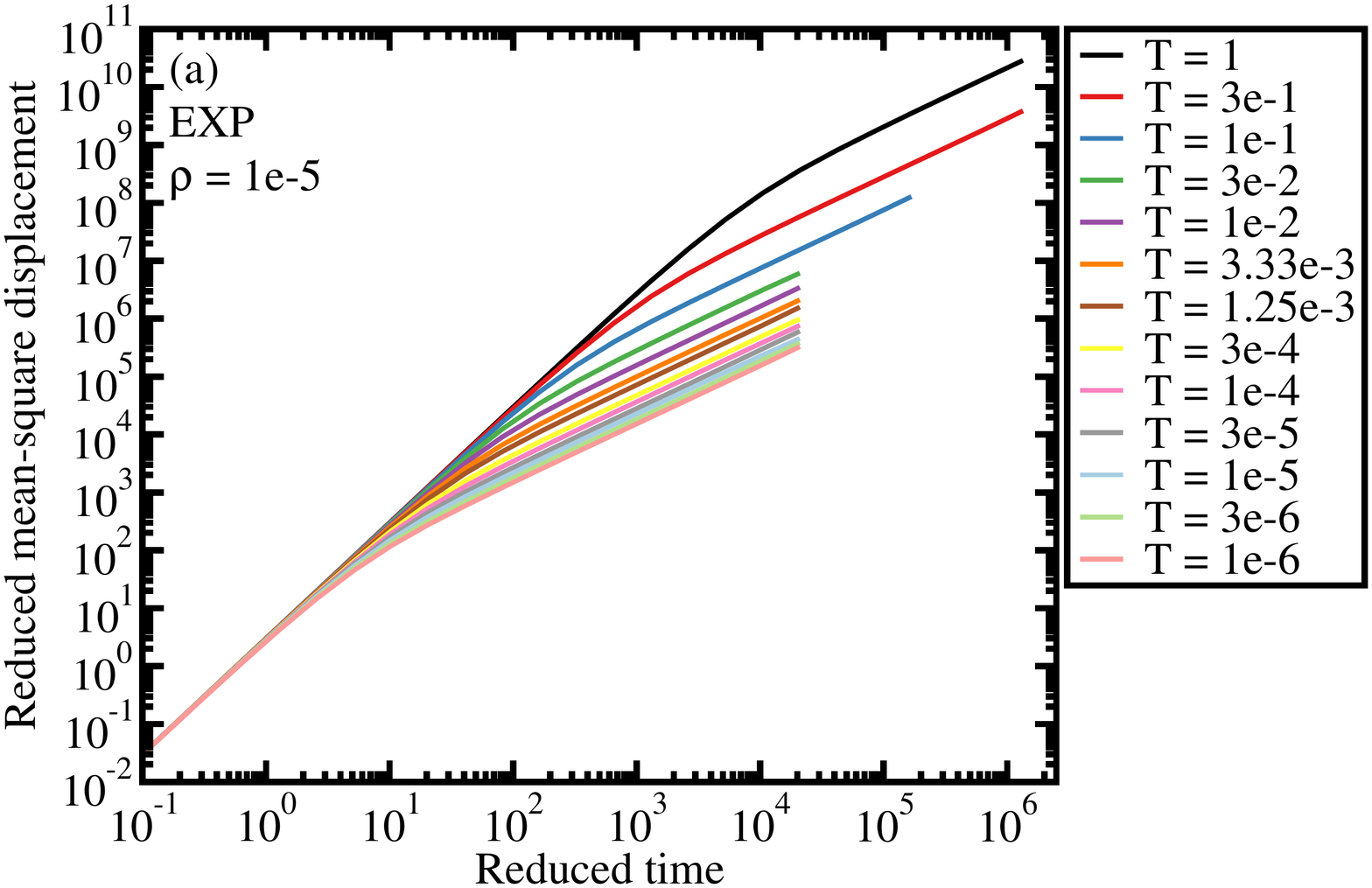}
	\includegraphics[width=0.45\textwidth]{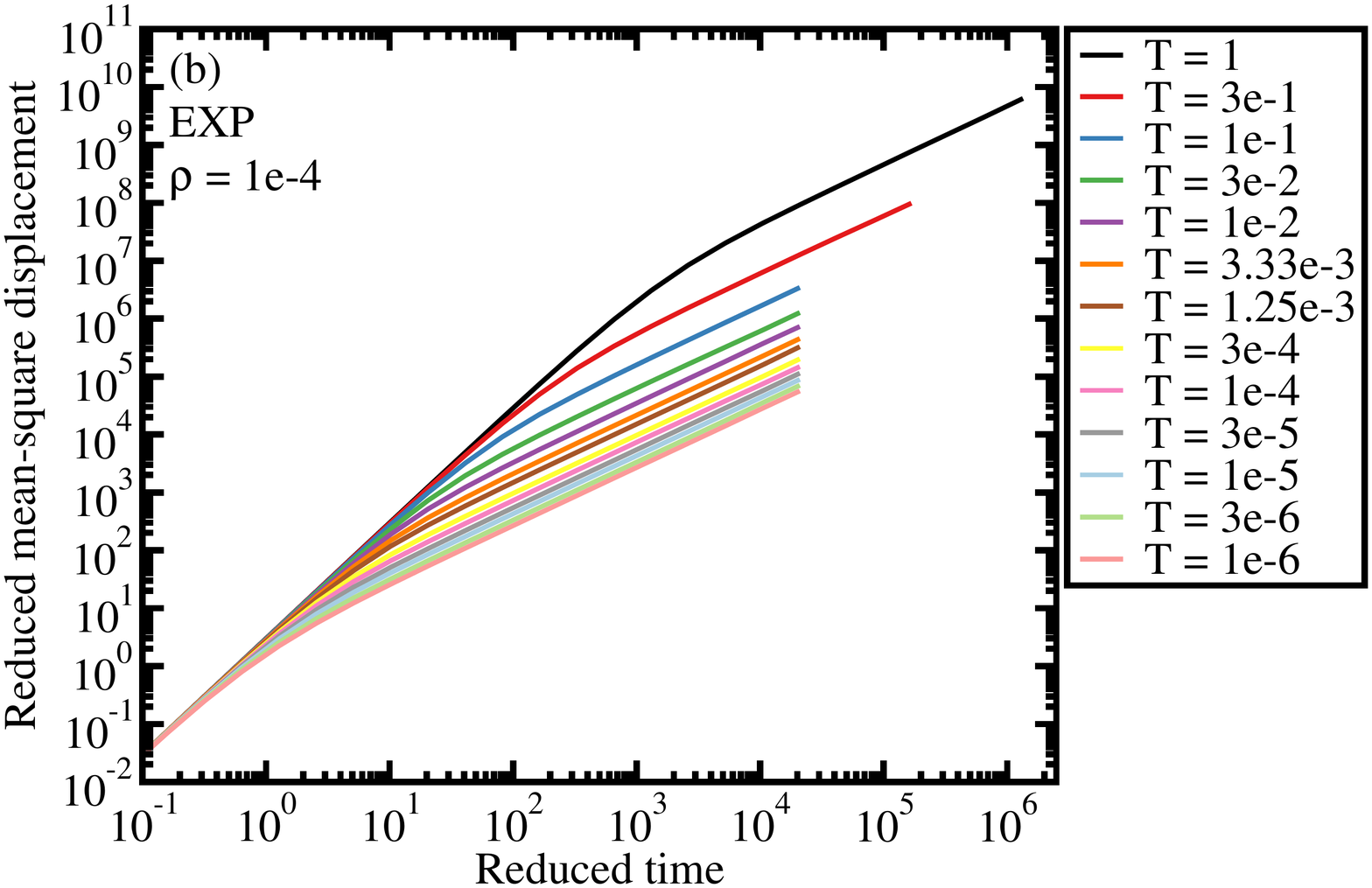}
	\includegraphics[width=0.45\textwidth]{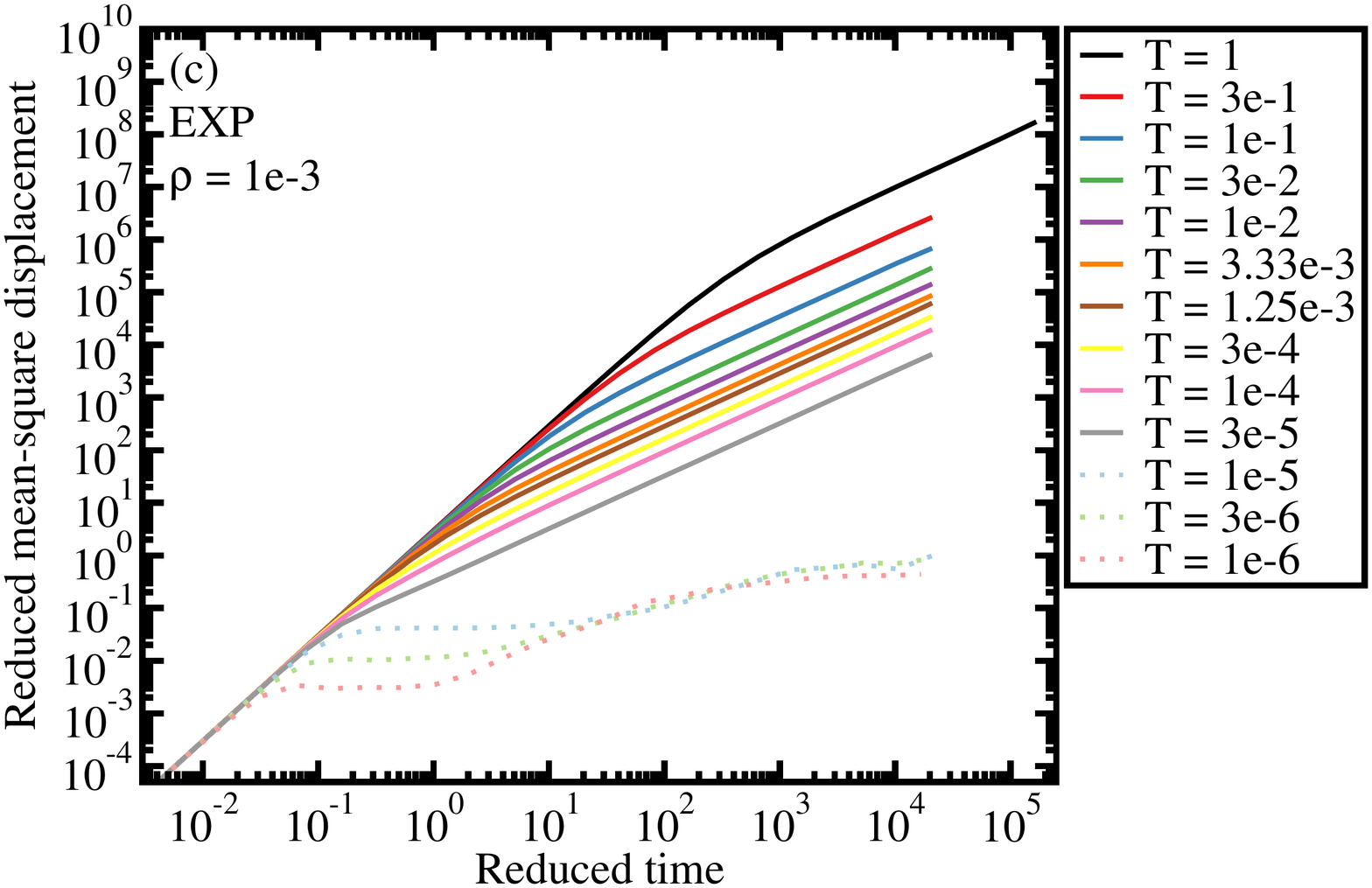}
	\includegraphics[width=0.45\textwidth]{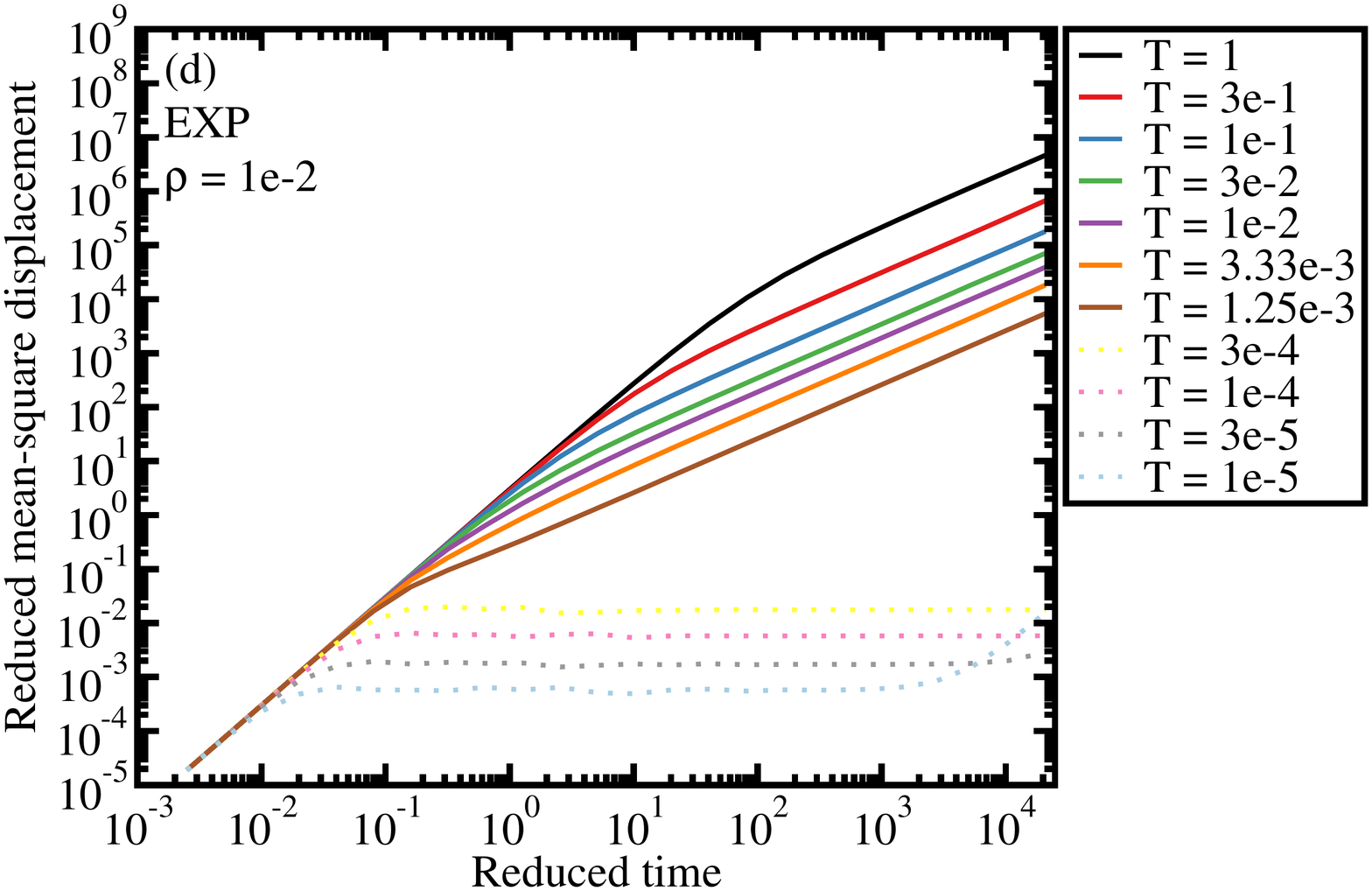}
	\caption{\label{fi:isor-msd}Reduced MSD along four isochores: (a) ${\rho} = 10^{-5}$, (b) ${\rho} = 10^{-4}$, (c)  ${\rho} = 10^{-3}$, (d) ${\rho} = 10^{-2}$. Gas and liquid phase state points are indicated by full lines, solid-phase state points by dotted lines. }
\end{figure}

	\begin{figure}[!htbp]
		\centering
		\includegraphics[width=0.45\textwidth]{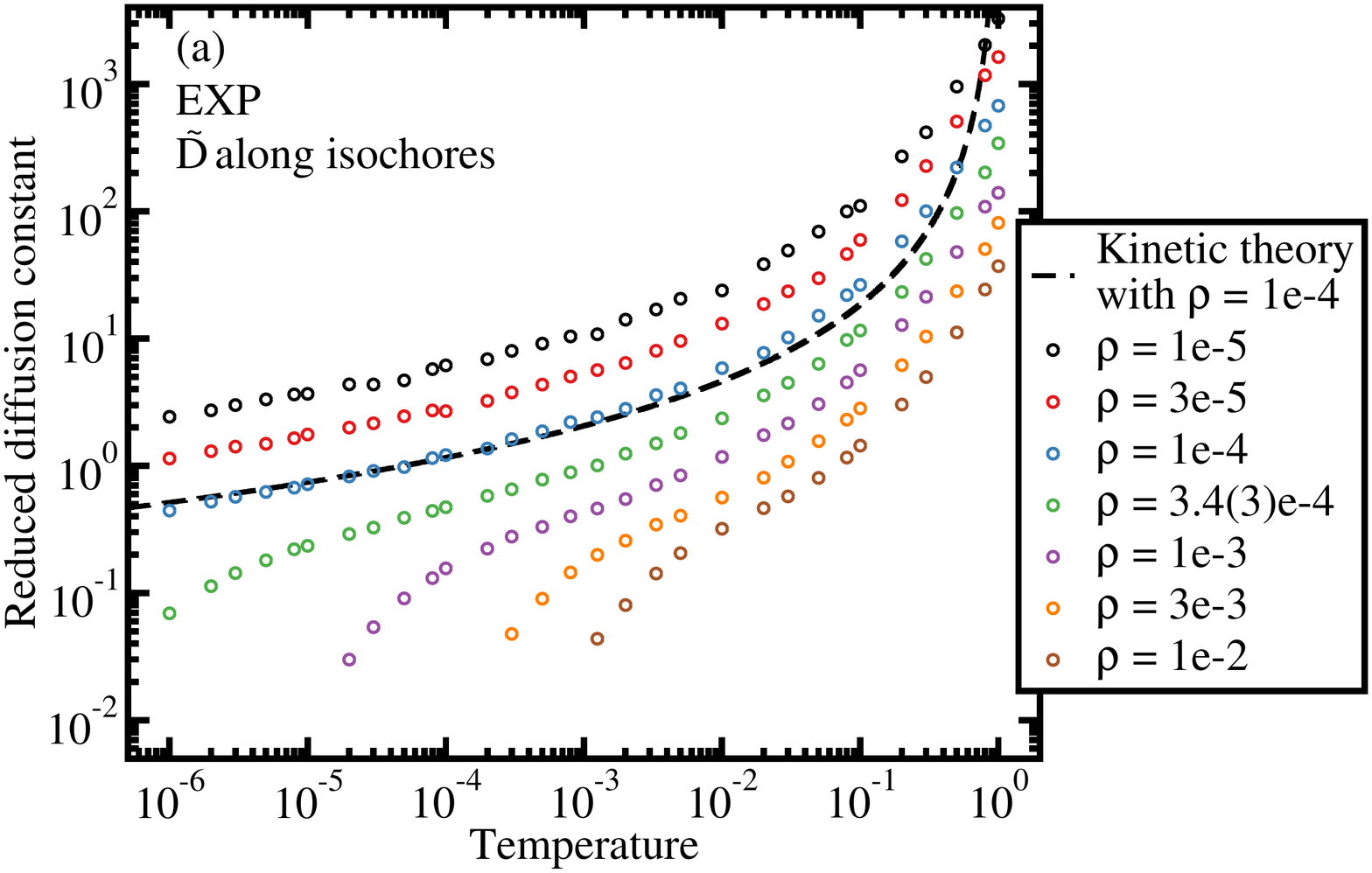}
		\includegraphics[width=0.45\textwidth]{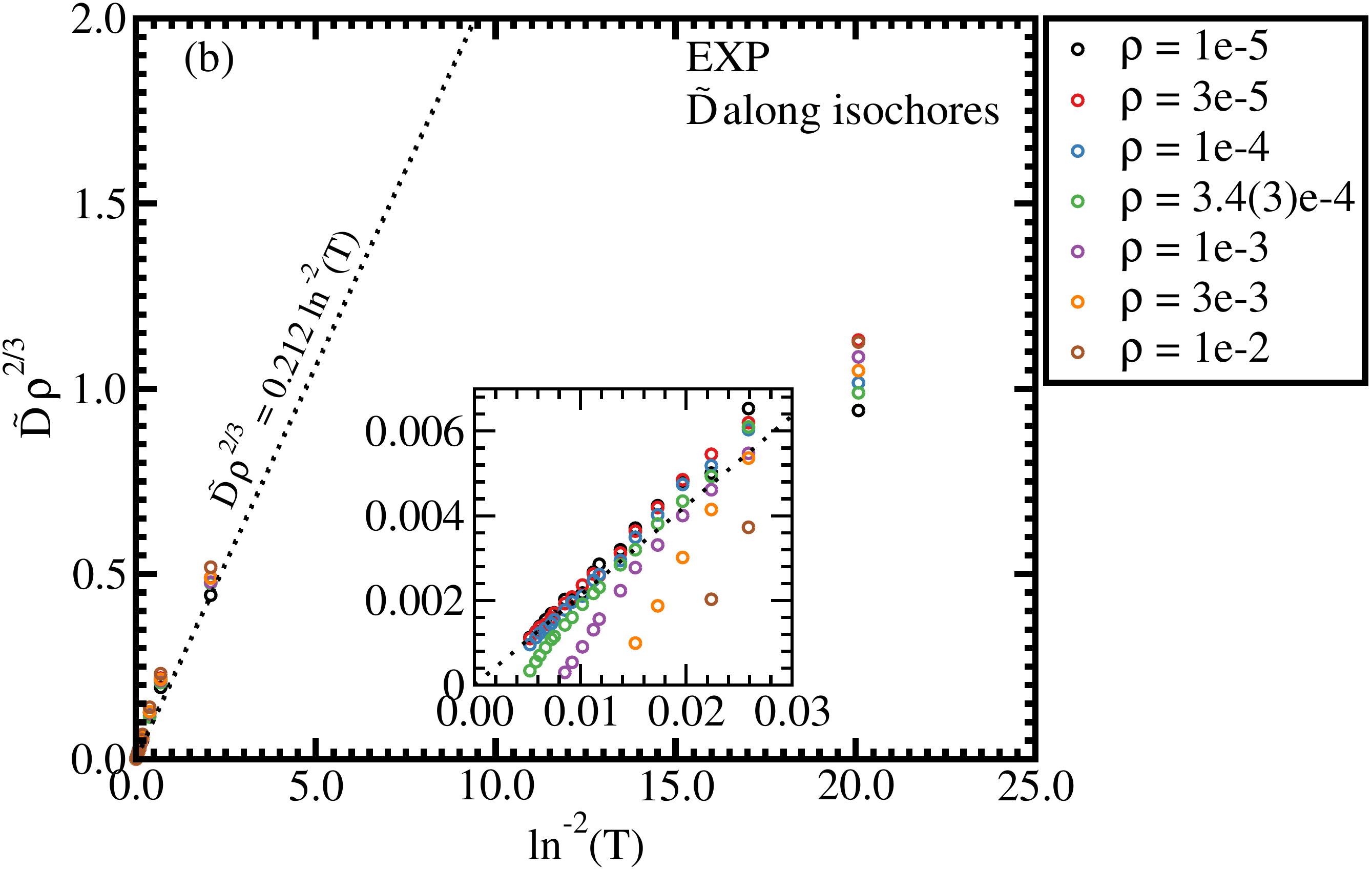}
		\caption{\label{fi:isor-D}
		(a) Reduced diffusion constants along seven isochores. The dashed line is the Enskog kinetic-theory prediction for $\rho=10^{-4}$, \eq{enskog}.  The inflection at low temperatures and high densities is where the condensed liquid phase is approached, signaling strong deviations from the gas-phase prediction. 			
		(b) investigates \eq{enskog} more closely by plotting $\rho^{2/3}\tilde{D}$ versus $1/\ln^2(T)$. The dotted line is the Enskog prediction, which works well at low densities. The inset focuses on the data close to the origin.}
		\end{figure}

\Fig{fi:isor-msd} shows the reduced MSD along the same isochores. At high temperatures the system is a dilute gas and the transition to diffusive behavior takes place much above $\tt\sim 1$, compare \fig{fi:isoT-msd}. 

\Fig{fi:isor-D}(a) shows how the reduced diffusion constant varies with temperature along seven isochores. For a given temperature the reduced diffusion constant is highest at low densities. The increase with temperature reflects the effective particle size decreasing, compare kinetic theory (Sec. \ref{IV}). The Enskog prediction \eq{enskog} for $\rho=10^{-4}$ is shown as the dashed line in \fig{fi:isor-D}(a). \Fig{fi:isor-D}(b) plots $\rho^{2/3}\tilde{D}$ versus $\ln^2(T)$ in order to test \eq{enskog}, which is expected to apply asymptotically as the density goes to zero. This is the case to a good approximation.

	\begin{figure}[!htbp]
		\centering
		\includegraphics[width=0.45\textwidth]{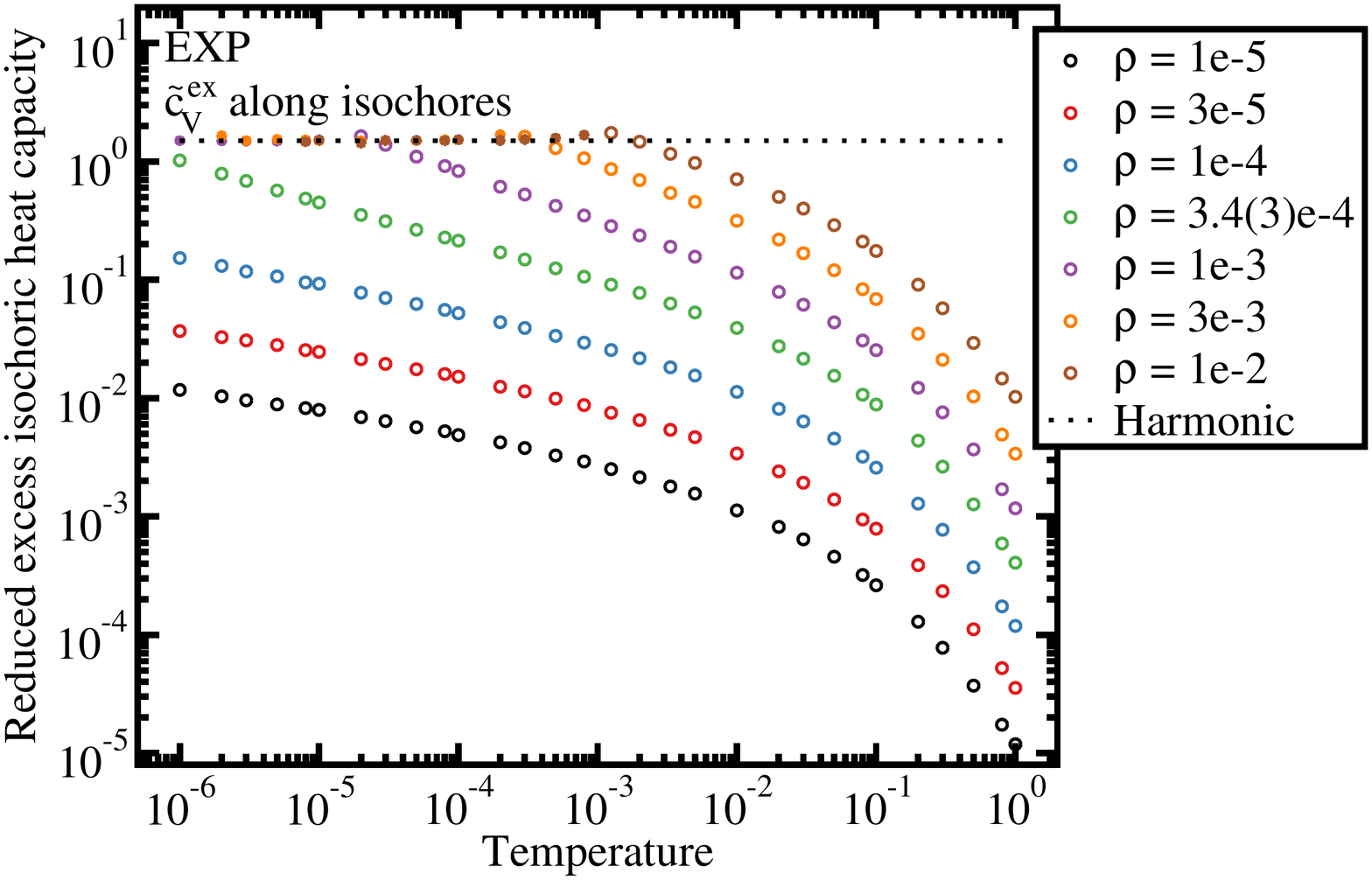}
		\caption{Reduced isochoric excess heat capacity per particle $\tcVex$ along seven isochores. Gas and liquid phase state points are represented by open symbol, stars correspond to solid-phase state points. The dashed line is the prediction of a harmonic crystal ($\tcVex=3/2$).}
		\label{fi:isor-C}
	\end{figure}

Finally, \fig{fi:isor-C} shows how the excess isochoric heat capacity $\tcVex$ varies with temperature along seven isochores. At low temperatures and high densities the reduced heat capacity is high and constant, close to the $3/2$ per particle harmonic contribution expected in the solid phase. For other state points the excess heat capacity is considerably lower.

\section{Quasiuniversality}\label{QUA}

As mentioned in Sec. \ref{I}, the EXP pair potential is central in a recent proof of simple liquids' quasiuniversality \cite{bac14a} (a brief review of which is given in Ref. \onlinecite{dyr16}). The idea is that -- to a good approximation -- any pair-potential system for which $v(r)$ is a sum of exponential functions taken from the strongly correlating part of the phase diagram (\fig{fi:phase-diagram}) has the same structure and dynamics as the EXP system itself. This section presents an $NVU$-based proof of quasiuniversality, combining arguments from Refs. \onlinecite{sch14} and \onlinecite{dyr13}. After this, as an example it is shown how the Lennard-Jones system's structure at four state points may be approximated by those of EXP pair-potential systems with the same reduced diffusion constant.

\subsection{$NVU$ proof of simple liquids' quasiuniversality}

$NVU$ dynamics is a molecular dynamics based on conservation of the potential energy \cite{NVU_I,NVU_II,NVU_III}. The idea is the following. The 3N-dimensional configuration space of the particle coordinates $\bR$ -- usually implemented assuming periodic boundary conditions, i.e., on a multidimensional torus -- has (3N-1)-dimensional hypersurfaces of constant potential energy. $NVU$ dynamics is defined as motion at constant velocity on these hypersurfaces along geodesic curves, the curves of (locally) minimum length. A geodesic curve is a  generalized straight line. $NVU$ dynamics may be regarded as realizing Newton's first law in the curved, high-dimensional space defined by a constant-potential-energy hypersurface. Geodesic dynamics also appears in the general theory of relativity, there just in four dimensions. Despite the fact that the potential and kinetic energies separately are conserved in $NVU$ dynamics, it has been shown analytically as well as numerically that $NVU$ dynamics in the thermodynamic limit leads to the same structure and dynamics as ordinary Newtonian $NVE$ or $NVT$ dynamics \cite{NVU_I,NVU_II}. 

Quasiuniversality of systems with a pair-potential function that is a sum of EXP pair potentials from the strongly correlating part of the EXP phase diagram (\fig{fi:phase-diagram}), is based on the following fact: For different pair-potential parameters $\varepsilon$ and $\sigma$, the EXP system's family of reduced-unit constant-potential-energy hypersurfaces are identical. We show this below, followed by a proof that systems with potential energy given as a linear combination of two or more EXP terms have the same constant-potential-energy hypersurfaces as the EXP system itself and, consequently, have the same $NVU$ trajectories. This implies identical structure and dynamics.

For any system at density $\rho$ one defines the microscopic excess entropy function by  $\Sex(\bR)\equiv\Sex(\rho,U)|_{U=U(\bR)}$ in which $\Sex(\rho,U)$ is the thermodynamic excess entropy as a function of density and average potential energy \cite{sch14}. In other words, $\Sex(\bR)$ is the thermodynamic excess entropy of the state point with density $\rho$ (corresponding to $\bR$) and average potential energy $U(\bR)$. By inversion one has $U(\bR)=U(\rho,\Sex(\bR))$ in which $U(\rho,\Sex)$ is the average potential energy at the state point with density $\rho$ and excess entropy $\Sex$. From the configuration-space microcanonical ensemble expression for the excess entropy it is straightforward to show that the hidden-scale-invariance condition \eq{hs} implies $\Sex(\lambda\bR)=\Sex(\bR)$, i.e., a uniform scaling of a configuration does not change its excess entropy \cite{sch14}. This means that the microscopic excess entropy is a function of the configuration's reduced coordinate vector $\tbR\equiv\rho^{1/3}\bR$ \cite{sch14}, implying that

\be\label{Rfundeq}
U(\bR)
\,=\, U(\rho,\Sex(\tbR)) \,.
\ee

Having in mind that the EXP system's potential-energy function $U_\esub(\bR)$ depends on $\varepsilon$ and $\sigma$, we note that \eq{Rfundeq} implies that a dimensionless function $\Phi_\esub$ of two variables exists such that ($\tSex\equiv\Sex/k_B$)

\be\label{EXPfundeq}
U_\esub(\bR,\varepsilon,\sigma)
\,=\, \varepsilon\,\Phi_\esub\left(\rho\sigma^3,\tSex^\esub(\tbR)\right)\,.
\ee
The appearances of $\varepsilon$ in front and of $\sigma^3$ multiplied by the density are dictated by dimensional analysis. A consequence of \eq{EXPfundeq} is that different EXP systems have the same family of reduced-coordinate constant-potential-energy hypersurfaces, which are given by $\tSex^\esub(\tbR)=\,$Const. 

Consider now the system defined by the pair potential $v(r)=\varepsilon_1\exp(-r/\sigma_1)+\varepsilon_2\exp(-r/\sigma_2)$ and let us focus on one particular configuration $\bR$. Since it defines all pair distances, by \eq{EXPfundeq} this system's potential-energy function is given by adding the two EXP system's potential energies,

\be\label{U_sum}
U(\bR)
\,=\,
 \varepsilon_1\,\Phi_\esub\left(\rho\sigma_1^3,\tSex^\esub(\tbR)\right)\,+\,\varepsilon_2\,\Phi_\esub\left(\rho\sigma_2^3,\tSex^\esub(\tbR)\right)\,.
\ee
Assuming positive temperature, i.e., that 

\be
\varepsilon_1(\partial\Phi_\esub(\rho\sigma_1^3,\Sex)/\partial\Sex)_\rho+\varepsilon_2(\partial\Phi_\esub(\rho\sigma_2^3,\Sex)/\partial\Sex)_\rho>0\,,
\ee 
the potential energy $U(\bR)$ of \eq{U_sum} can be constant only if $\tSex^\esub(\tbR)$ is constant. Via \eq{EXPfundeq} this implies that $U_\esub(\bR,\varepsilon,\sigma)$ is constant. Thus the constant-potential-energy hypersurfaces for the function $U(\bR)$ of \eq{U_sum} are -- except for a uniform scaling -- identical to the EXP system's constant-potential-energy hypersurfaces. The reduced-unit $NVU$ dynamics is consequently identical to that of the EXP system, implying identical structure and dynamics.

The above generalizes to pair-potential systems of arbitrary linear combinations of EXP terms, and EXP pair-potential terms may also be subtracted. Basically, the only requirement is that each EXP pair-potential term refers to the strongly correlating part of the EXP systems phase diagram, compare \fig{fi:phase-diagram} \cite{dyr16}. This requirement, which ensures that \eq{EXPfundeq} applies for each term, translates into requiring that the reduced-unit pair potential in question is a sum of EXP terms with numerically large prefactors \cite{dyr16}.

\subsection{Example: The EXP system approximates the Lennard-Jones system}

The Lennard-Jones (LJ) system ($v(r)=4\varepsilon\left[(r/\sigma)^{-12}-(r/\sigma)^{-6}\right]$) is quasiuniversal \cite{han13,bac14a}. As a demonstration of quasiuniversality we consider four state points of the LJ system typical for the high-temperature gas, the high-temperature liquid, and the liquid close to the melting line. At each state point the reduced diffusion constant $\tilde{D}$ was evaluated. According to quasiuniversality $\tilde{D}$ determines the reduced structure. For each of the four reduced LJ diffusion constants we identified two or three EXP systems (equivalently: EXP state points) with the same $\tilde{D}$ and calculated the RDF in order to compare to those of the LJ system. 

\begin{figure}[!htbp]
	\centering
	\includegraphics[width=0.45\textwidth]{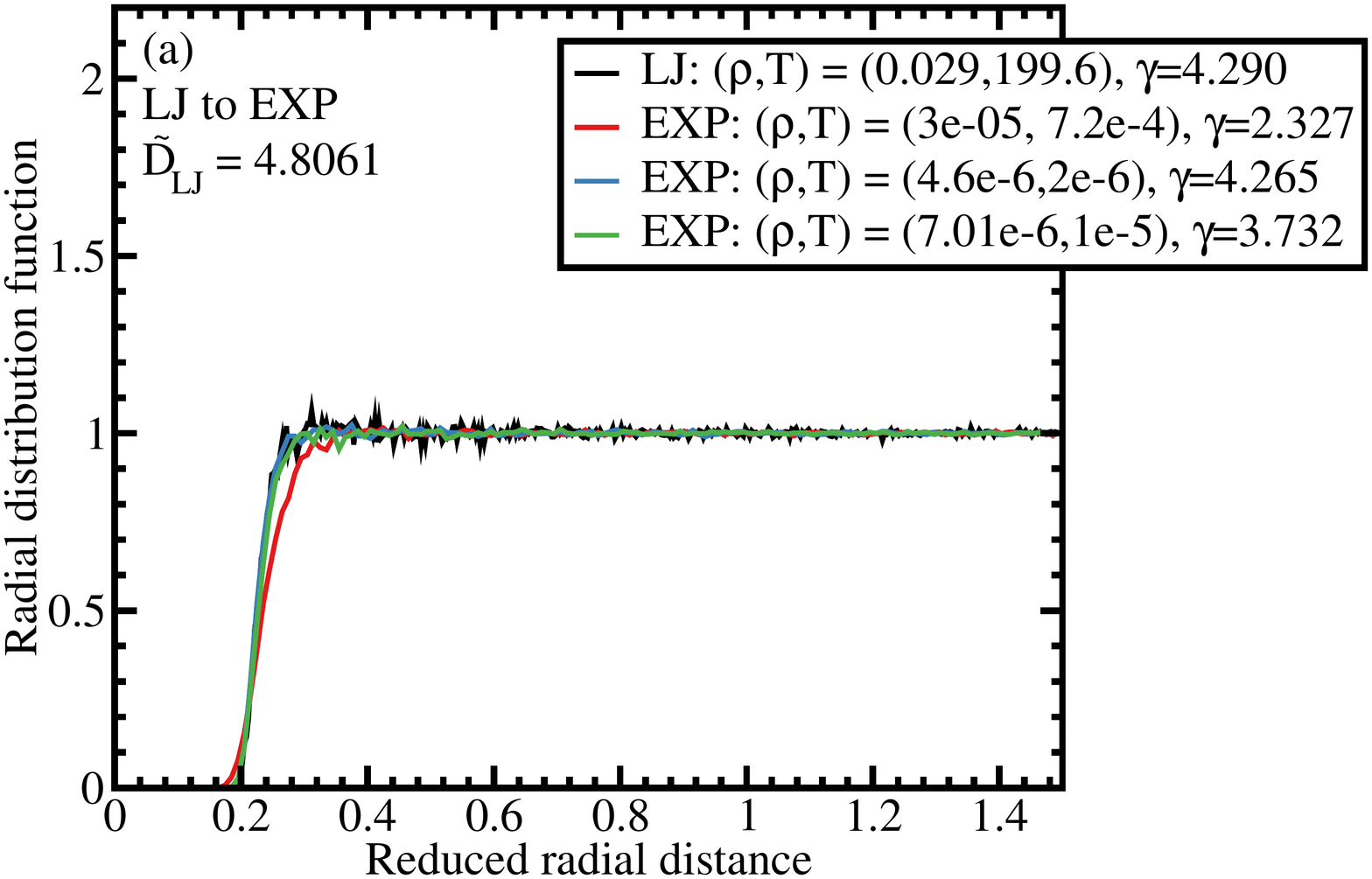}
	\includegraphics[width=0.45\textwidth]{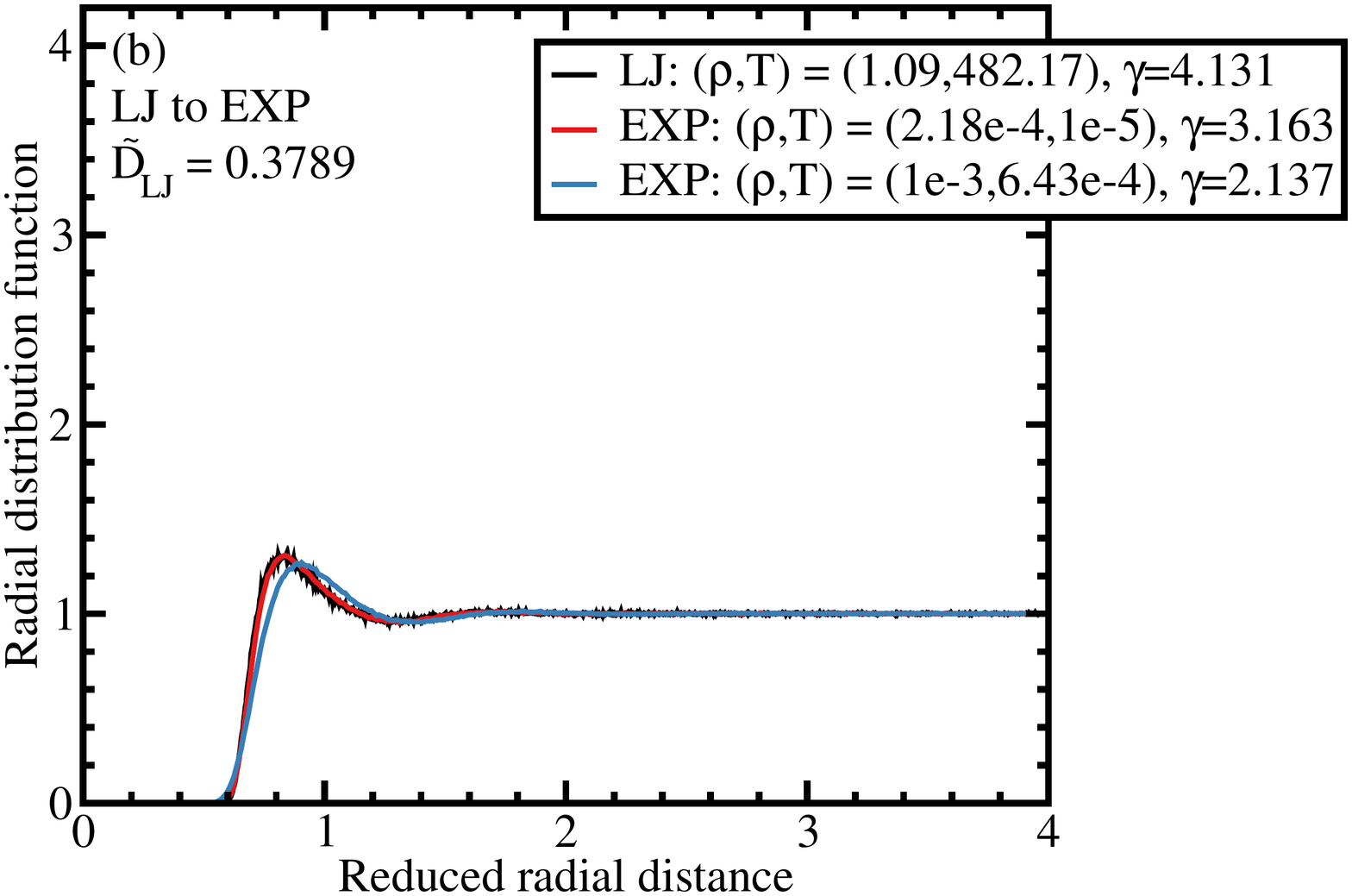}	
	\includegraphics[width=0.45\textwidth]{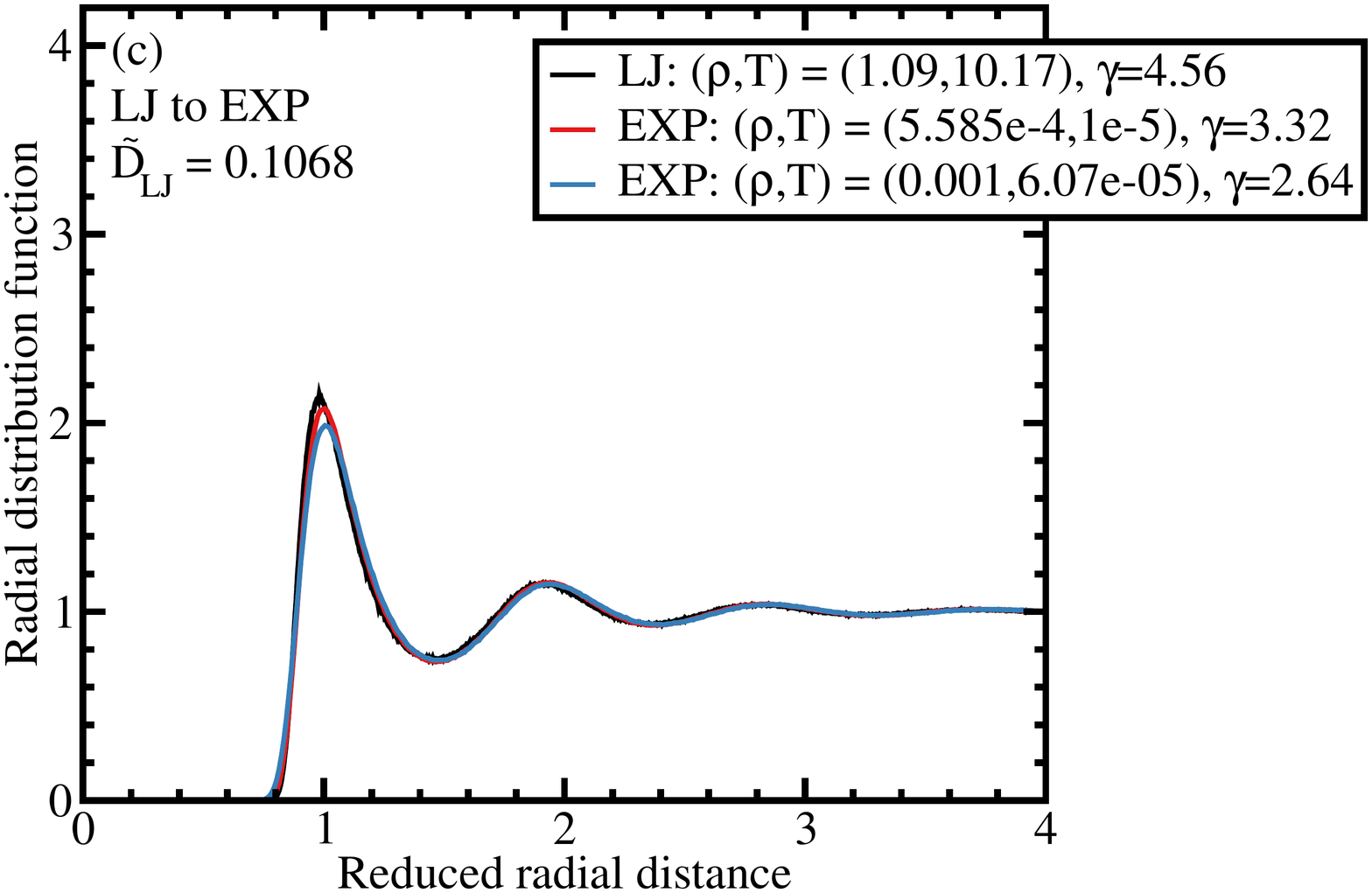}	
	\includegraphics[width=0.45\textwidth]{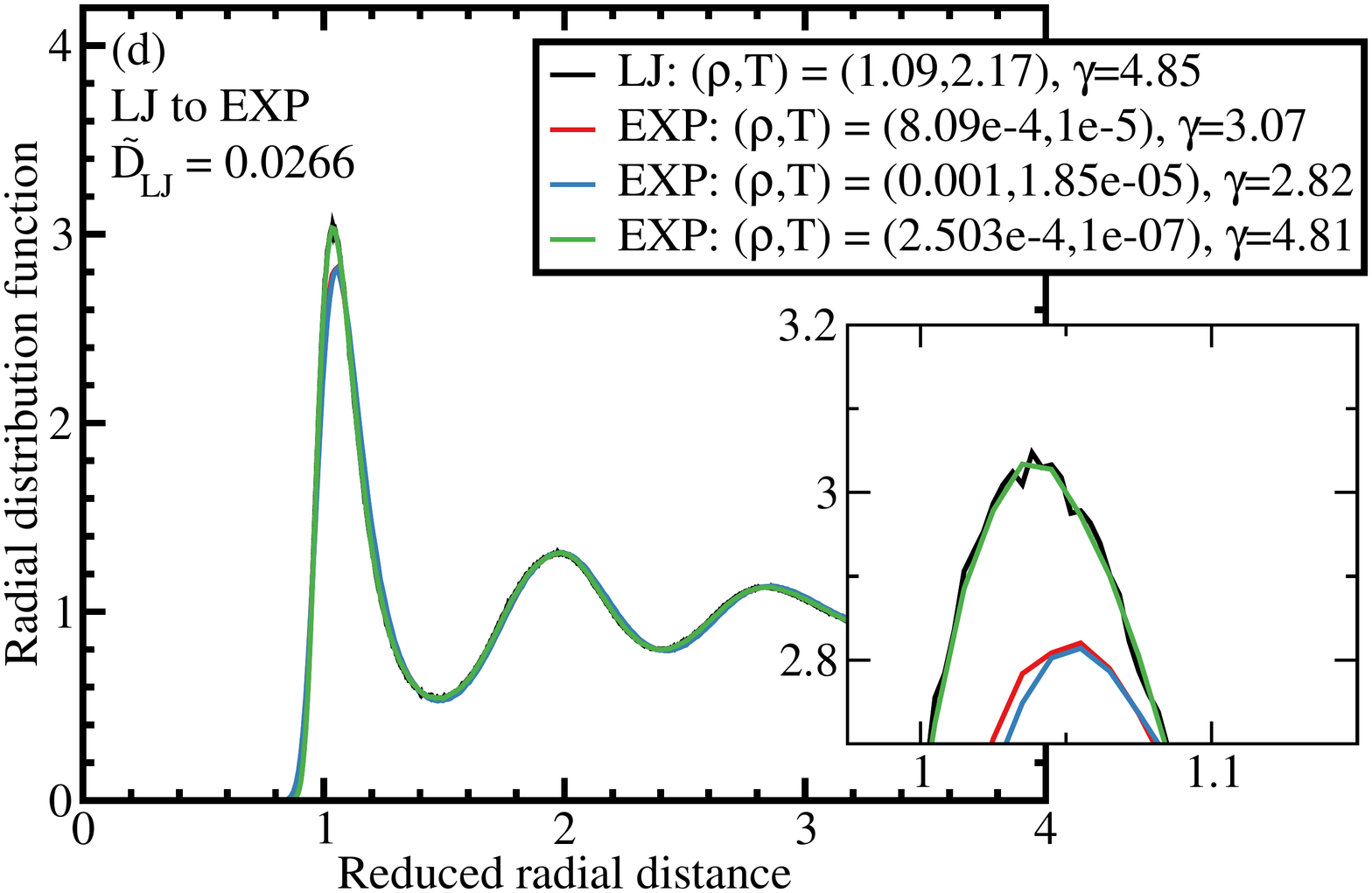}
	\caption{Quasiuniversality illustrated by comparing radial distribution functions (RDF) at four state points for the Lennard-Jones (LJ) system (black curves) to those of EXP systems with the same reduced diffusion constant (within 1\%, colored curves). EXP state points are specified by density, temperature, and density-scaling exponent $\gamma$.
	(a) LJ state point $(\rho,T)=(0.029,199.6)$, a typical high-temperature gas state point at which $\tilde{D}=4.8061$. $\gamma$ is here $4.29$, which is not far from the value $4$ predicted from the repulsive $r^{-12}$ term of the LJ pair potential \cite{II}. The red, blue and green curves are RDF predictions for different EXP systems with the same reduced diffusion constant.
	(b) LJ state point $(\rho,T)=(1.09,482.17)$, a moderate-density, high-temperature gas state point at which $\tilde{D}=0.3789$. The red EXP system fits better than the blue one. Deviations are centered around the first peak, with largest deviations for the EXP state point with density-scaling exponent $\gamma$ most different from its LJ value (blue). 
	(c) LJ state point $(\rho,T)=(1.09,10.17)$ at which $\tilde{D}=0.1068$. There are slight deviations around the first peak; again these are smallest for the EXP system with $\gamma$ closest to that of the LJ system.
	(d) LJ state point $(\rho,T)=(1.09,2.17)$, a condensed-phase liquid state point close to the melting line at which $\tilde{D}=0.0266$. The green curve, which fits best, represents an EXP system that has virtually the same $\gamma$ as the LJ system. The inset provides a blow up of the first peak.	 \label{kvasfig}}
\end{figure}

The results are shown in \fig{kvasfig}, which gives LJ system RDFs as black curves and those of EXP systems with same reduced diffusion constant as colored curves. The fits are generally good. Deviations center around the first peak. These reflect the following breakdown of quasiuniversality: At small interparticle separations the RDF is dominated by the pair potential via the asymptotic behavior $g(r)\sim\exp(-v(r)/k_BT)$ for $r\rightarrow 0$ \cite{han13}. The quantity $v(r)/k_BT$ is not isomorph invariant, however, implying that the way in which $g(r)$ approaches zero at short distances must violate quasiuniversality. If one assumes that the number of particles in the first coordination shell \textit{is} quasiuniversal, there must be a non-quasiuniversal signature on the height of the first peak of the RDF. For the EXP system, the larger the density-scaling exponent becomes along an isomorph (see Paper II) the higher is the peak because the more rapidly does $g(r)$ go to zero at short distances. This explains the slight deviations from quasiuniversality observed in \fig{kvasfig}. If one wishes from the reduced diffusion constant to identify an EXP system with almost identical RDF also around the first peak, an EXP system should be sought with both correct reduced diffusion constant and correct density-scaling exponent. This is illustrated in \fig{kvasfig}(d), compare the inset.

\section{Concluding remarks}\label{VI}

We have presented an investigation of the EXP pair-potential system's structure and dynamics over a large part of its low-temperature, low-density thermodynamic phase diagram, focusing on gas and liquid state points. At temperatures higher than those studied here the EXP system changes character because particles may overlap and pass through one another. As for other systems with no attractive forces, the EXP system has a solid and a fluid phase, but no liquid-gas phase transition. We find gas-like behavior in a large part of the studied phase diagram as revealed by a  virtual absence of structure probed by the RDF. When varying density or temperature we find, not surprisingly, that the EXP fluid has more structure the closer it is to the melting transition. For both structure and dynamics one finds the same trends whether density or temperature is lowered. This reflects the existence of isomorphs, which are lines in the thermodynamic phase diagram along which the reduced-unit physics is invariant \cite{IV,sch14} (Paper II).

The motivation for studying the EXP pair-potential system is the recent suggestion that the EXP potential may be regarded as ``the mother of all pair potentials'' in the sense that any pair potential of an R-simple single-component pair-potential system may be well approximated by a sum of EXP pair potentials with coefficient that in reduced units are numerically much larger than unity \cite{bac14a,dyr16}. Because the EXP system is R-simple, isomorph theory implies that any such linear combination has virtually the same structure and dynamics as the pristine EXP system \cite{dyr16}, compare Sec. \ref{QUA}. This is our explanation of the quasiuniversality that has been reported for the majority of simple liquids and which is traditionally explained by reference to the hard-sphere system.

The present paper focused on the gas and liquid phases. This is also the focus of the companion paper studying the EXP system's isomorphs \cite{EXP_II_arXiv}.

\acknowledgments{We thank Lorenzo Costigliola for helpful discussions. This work was supported by the VILLUM Foundation's \textit{Matter} grant (16515).}

\section*{APPENDIX I: Analytical theory for the virial potential-energy correlation coefficient in the gas phase}

For mathematical simplicity we use below the EXP unit system in which $\varepsilon=\sigma=1$; moreover we put $k_B=1$. The EXP pair potential is given by

\be\label{EXPpp}
v(r)\,=\, e^{-r}\,.
\ee
The inverse temperature is denoted by $\beta$, i.e., $\beta\equiv 1/T$. 

When the density is sufficiently low, the individual pair energies and forces are statistically independent and one can calculate the averages in $R$ (\eq{R_def}) by reference to single particle pairs. For such a pair at distance $r$ the virial is given by $w=(-1/3)rv'(r)$, i.e.,

\be
w
\,=\,\frac{1}{3}\ln(1/v)v\,.
\ee
In terms of $v$ and $w$ \eq{R_def} becomes

\be\label{Rsq}
R
\,=\,\frac{\langle vw\rangle-\langle v\rangle\langle w\rangle}{\sqrt{(\langle v^2\rangle-\langle v\rangle^2)(\langle w^2\rangle-\langle w\rangle^2)}}\,.
\ee

The gas-phase physics is determined by the two-particle Boltzmann canonical probability, $p(r)\propto r^2\exp(-\beta v(r))$. The pair-potential energy $v=\exp(-r)$ varies between zero and one, but when the temperature is low, little error arises from allowing $v$ to be any positive number. The probability of finding the pair potential energy $v$ is given by $p(v)=p(r)|dr/dv|$, i.e., since $r=\ln(1/v)$ one has $p(v)\propto\ln^2(1/v)\exp(-\beta v)\left|{dr}/{dv}\right|$ or

\be\label{pv2}
p(v)
\,\propto\,\frac{\ln^2(1/v)\exp(-\beta v)}{v}\,\,\,\,\,(0<v<\infty)\,.
\ee
This distribution is not normalizable in the $v\rightarrow 0$ limit, reflecting the infinitely many particle pairs found far from each other. Introducing a lower $v$ cut-off, the normalization constant thus diverges as the cutoff goes to zero. This means that in expressions like $\langle\Delta v\Delta w\rangle=\langle v w\rangle-\langle v\rangle\langle w\rangle$ the latter product disappears as the $v$ cutoff goes to zero, so \eq{Rsq} simplifies into

\be
R
\,=\,\frac{\langle v w\rangle}{\sqrt{\langle v^2\rangle\langle w^2\rangle}}\,.
\ee
If one defines

\be
A_n
\,=\,\int_{0}^{\infty} v\,\ln^n(1/v)\,e^{-\beta v}dv\,
\ee
and $K$ is the normalization constant of $p(v)$, one has $\langle v^2\rangle=KA_2$, $\langle vw\rangle=KA_3/3$, and $\langle w^2\rangle=KA_4/9$, but fortunately $K$ does not enter into the final expression,

\be
R
\,=\,\frac{A_3}{\sqrt{A_2\,A_4}}\,.
\ee

Using Maple one gets

\be
A_2
\,=\,\beta^{-2}\Big(
\ln^2\beta-2(1-C)\ln\beta+(\pi^2/6+C^2-2C)\Big)\,,
\ee

\be
A_3
\,=\,\beta^{-2}\Big(
\ln^3\beta-3(1-C)\ln^2\beta+(\pi^2/2+3C^2-6C)\ln\beta+k_3\Big)\,,
\ee
and

\be
A_4
\,=\,\beta^{-2}\Big(
\ln^4\beta-4(1-C)\ln^3\beta+(\pi^2+6C^2-12C)\ln^2\beta+h_4\ln\beta+k_4\Big)\,.
\ee
Here

\be
\Eu\,\equiv\lim\limits_{n\rightarrow\infty}\left(\sum_{p=1}^{n}\frac{1}{p}\,-\ln n\right)\,=\,0.577216...\,
\ee
is Euler's constant (in his original notation, this number is sometimes denoted by $\gamma$),

\be
k_3
\,=\,C^3-3C^2+(\pi^2/2)C-\pi^2/2+2\zeta(3)
\,=\,-0.48946\,
\ee
in which $\zeta(3)=1.20206$ is the Riemann zeta function's value at $3$ (``Apery's constant''), 

\be
h_4
\,=\,2\left(2C^3-6C^2+\pi^2C-\pi^2+4\zeta(3)\right)
\,=\,4k_3
\,=\,-1.9578\,,
\ee
and

\be
k_4
\,=\,C^4-4C^3+\pi^2C^2-2\pi^2C+8\zeta(3)C-8\zeta(3)+3\pi^4/20
\,=\,1.7820\,.
\ee

Numerically, the three integrals are given by

\be
\,\beta^2\,A_2
\,=\,\ln^2\beta-0.8456\ln\beta+0.8237\,,
\ee

\be
\,\beta^2\,A_3
\,=\,
\ln^3\beta-1.268\ln^2\beta+2.471\ln\beta-0.4895\,,
\ee
and

\be
\,\beta^2\,A_4
\,=\,
\ln^4\beta-1.691\ln^3\beta+4.942\ln^2\beta-1.958\ln\beta+1.782\,.
\ee

\section*{Appendix II: Simulated state points}

The state points simulated involve the following densities: 
$1.00\cdot 10^{-5}; 2.00\cdot 10^{-5}; 3.00\cdot 10^{-5}; 5.00\cdot 10^{-5}; 8.00\cdot 10^{-5}; 
 1.00\cdot 10^{-4}; 1.25\cdot 10^{-4}; 2.16\cdot 10^{-4}; 3.43\cdot 10^{-4}; 5.12\cdot 10^{-4}; 7.29\cdot 10^{-4}; 
 1.00\cdot 10^{-3}; 2.00\cdot 10^{-3}; 3.00\cdot 10^{-3}; 5.00\cdot 10^{-3}; 8.00\cdot 10^{-3}; 
 1.00\cdot 10^{-2}$; and the following temperatures:
$1.00\cdot 10^{-6}; 2.00\cdot 10^{-6}; 3.00\cdot 10^{-6}; 5.00\cdot 10^{-6}; 8.00\cdot 10^{-6};
 1.00\cdot 10^{-5}; 2.00\cdot 10^{-5}; 3.00\cdot 10^{-5}; 5.00\cdot 10^{-5}; 8.00\cdot 10^{-5}; 
 1.00\cdot 10^{-4}; 2.00\cdot 10^{-4}; 3.00\cdot 10^{-4}; 5.00\cdot 10^{-4}; 8.00\cdot 10^{-4}; 
 1.25\cdot 10^{-3}; 2.00\cdot 10^{-3}; 3.33\cdot 10^{-3}; 5.00\cdot 10^{-3}; 
 1.00\cdot 10^{-2}; 2.00\cdot 10^{-2}; 3.00\cdot 10^{-2}; 5.00\cdot 10^{-2}; 8.00\cdot 10^{-2}; 
 1.00\cdot 10^{-1}; 2.00\cdot 10^{-1}; 3.00\cdot 10^{-1}; 5.00\cdot 10^{-1}; 8.00\cdot 10^{-1};
 1$.

\end{document}